\def\inarxiv{}
\documentclass[10pt,conference]{IEEEtran} %
\IEEEoverridecommandlockouts

\def\BibTeX{{\rm B\kern-.05em{\sc i\kern-.025em b}\kern-.08em
T\kern-.1667em\lower.7ex\hbox{E}\kern-.125emX}}

\PassOptionsToPackage{hyphens}{url}
\usepackage[hyphens]{url}
\PassOptionsToPackage{breaklinks,colorlinks}{hyperref}
\usepackage[breaklinks,colorlinks]{hyperref}
\PassOptionsToPackage{dvipsnames}{xcolor}
\usepackage[dvipsnames]{xcolor}
\hypersetup{
  colorlinks,
  linkcolor={red!50!black},
  citecolor={blue!50!black},
  urlcolor={blue!50!black}
}
\usepackage{amsmath,amsopn}
\usepackage{colortbl}
\usepackage{endnotes,microtype,xspace,graphicx,fancyvrb,multirow}
\usepackage{booktabs}
\usepackage{array,underscore,relsize}
\usepackage[T1]{fontenc}
\usepackage{times}
\usepackage{fancyhdr}
\usepackage[inline]{enumitem}
\usepackage[labelfont=bf,font=small,skip=5pt]{caption}
\usepackage[belowskip=0pt,aboveskip=2pt]{subcaption}
\usepackage[normalem]{ulem}

\usepackage{fp}
\usepackage{siunitx}

\makeatletter
\@namedef{ver@lineno.sty}{9999/12/31}
\@namedef{opt@lineno.sty}{}
\makeatother

\usepackage{minted} %

\usepackage{algorithm}
\usepackage[noend]{algpseudocode}
\algrenewcommand\algorithmicindent{0.75em}

\usepackage{balance}
\usepackage{multicol}

\usepackage[super]{nth}
\usepackage{pifont}
\usepackage{import}
\usepackage{placeins}
\usepackage{longtable}
\usepackage{supertabular}
\usepackage{xr}
\usepackage{pgffor}
\usepackage{comment}
\usepackage{mdframed}
\usepackage{float}

\usepackage{fancyvrb}

\newcommand{\sys}{\mbox{\textsc{Msg}}\xspace}
\newcommand{\sysinline}{$\sys_{\text{inline}}$\xspace}
\newcommand{\sysm}{$\sys^{-}$\xspace}
\newcommand{\msg}{\mbox{\textsc{Msg}}\xspace} %
\newcommand{\autoverus}{\mbox{\textsc{AutoVerus}}\xspace} %

\newcommand{\sysaio}{$\sys_{\text{AIO}}$\xspace}
\newcommand{\sysaionaive}{$\sys_{\text{AIO-naive}}$\xspace}
\newcommand{\sysaiom}{$\sys_{\text{AIO}}^{-}$\xspace}
\newcommand{\sysaioinline}{$\sys_{\text{AIO-inline}}$\xspace}

\newif\ifrev
\ifdefined\inrev
  \revtrue
  \usepackage[
  draft,
  commandnameprefix=ifneeded,
  authormarkup=none,
  commentmarkup=margin]{changes}
  
  \newcommand{\revd}[1]{\textcolor{red}{\sout{#1}}} %
  \newcommand{\revonly}[1]{\textcolor{blue}{#1}}
\else
  \revfalse
  \usepackage[final,commandnameprefix=ifneeded]{changes}
  
  \newcommand{\revd}[1]{\ignorespaces}
  \newcommand{\revonly}[1]{\ignorespaces} %
\fi

\newif\ifarxiv
\ifdefined\inarxiv
  \arxivtrue
  \newcommand{\arxivonly}[1]{#1}
  \newcommand{\mainonly}[1]{\ignorespaces}
  \newcommand{\mainarxiv}[2]{#2}
  \usepackage{manyfoot}
  \makeatletter
  \def\arxivfootnoterule{\kern-3\p@
    \hrule \@width \columnwidth \kern 2.6\p@}
  \makeatother
  \SelectFootnoteRule[2]{arxiv}
  \DeclareNewFootnote[plain]{arxiv}
  
\else
  \arxivfalse
  \newcommand{\arxivonly}[1]{\ignorespaces}
  \newcommand{\mainonly}[1]{#1}
  \newcommand{\mainarxiv}[2]{#1}
\fi

\NewDocumentCommand{\TODO}{s m}{%
  \IfBooleanTF{#1}{}{%
    \noindent{\color{Melon} {\bf \fbox{TODO:}} #2}%
  }%
}

\definecolor{darkgreen}{RGB}{0,128,0}

\newcommand{\cc}[1]{\mbox{\smaller[0.5]\texttt{#1}}}

\fvset{fontsize=\scriptsize,xleftmargin=8pt,numbers=left,numbersep=5pt}

\input{code/fmt}
\IfFileExists{code/fmt.tex}
{\input{code/fmt}}
{\input{code/fmt-default}}

\def\Snospace~{\S{}}

\newif\ifdraft\drafttrue
\newif\ifnotes\notestrue
\ifdraft\else\notesfalse\fi

\input{glyphtounicode}
\newcolumntype{R}[1]{>{\raggedleft\let\newline\\\arraybackslash\hspace{0pt}}p{#1}}

\newcommand{\includepdf}[1]{
  \includegraphics[width=\columnwidth]{#1}
}

\newcommand{\squishlist}{
  \begin{itemize}[noitemsep,nolistsep]
      \setlength{\itemsep}{-0pt}
    }
    \newcommand{\squishend}{
  \end{itemize}
}

\newenvironment{itemizesqn}[1][]{
  \begin{itemize}[noitemsep,nolistsep,leftmargin=*,#1]
      \setlength{\itemsep}{-0pt}
    }{
  \end{itemize}
}

\newenvironment{enumeratesqn}[1][]{
  \begin{enumerate}[noitemsep,nolistsep,leftmargin=*,#1]
      \setlength{\itemsep}{-0pt}
    }{
  \end{enumerate}
}

\usepackage{tikz}
\newcommand*\WC[1]{%
  \begin{tikzpicture}[baseline=(C.base)]
    \node[draw,circle,inner sep=0.2pt](C) {#1};
\end{tikzpicture}}

\newcommand*\BC[1]{%
  \begin{tikzpicture}[baseline=(C.base)]
    \node[draw,circle,fill=black,inner sep=0.2pt](C) {\textcolor{white}{#1}};
\end{tikzpicture}}

\usepackage{xstring}
\newcommand{\PP}[1]{
  \vspace{2px}
  \noindent{\bf \IfEndWith{#1}{.}{#1}{#1.}}
}

\newcommand{\PN}[1]{
  \vspace{2px}
  \noindent{\bf #1}
}

\newcommand{\PPI}[1]{
  \noindent{\bf \IfEndWith{#1}{.}{#1}{#1.}}
}

\newcommand{\ie}{\textit{i}.\textit{e}.}
\newcommand{\eg}{\textit{e}.\textit{g}.}

\newcommand{\boxbeg}{
  \vspace{2px}
  \noindent
  \begin{tabular}{|l|}\hline
    \begin{minipage}{3.2in}
      \vspace{2px}
      \noindent
    }

    \newcommand{\boxend}{
      \vspace{2px}
    \end{minipage}\\ \hline
  \end{tabular}
  \vspace{-10pt}
}

\definecolor{ForestGreen}{RGB}{34,139,34}
\newcommand{\cmark}{\color{ForestGreen}\ding{51}}
\newcommand{\xmark}{\color{red}\ding{55}}

\makeatletter
\newcommand*{\addFileDependency}[1]{%
  \typeout{(#1)}%
  \@addtofilelist{#1}
  \IfFileExists{#1}{}{\typeout{No file #1.}}
}\makeatother

\newcommand{\replacex}[1]{%
  \StrLen{#1}[\stringlength]%
  \newcount\loopcounter
  \loopcounter=0
  \loop\ifnum\loopcounter<\stringlength%
  x%
  \advance\loopcounter by 1%
  \repeat%
}

\definecolor{codegray}{rgb}{0.5,0.5,0.5}
\definecolor{backgray}{rgb}{0.95,0.95,0.95}
\definecolor{redhighlight}{rgb}{1,0.6,0.6}

\usepackage{refcount} %

\NewDocumentCommand{\autorefp}{s m}{%
  \IfBooleanTF{#1}{%
    \autoref*{#2}%
  }{%
    \autoref{#2}%
  }%
  \ifnum\getpagerefnumber{#2}=\thepage
  \else
    \IfBooleanTF{#1}{%
    \ (page~\pageref*{#2})%
    }{%
    \ (page~\pageref{#2})%
    }%
  \fi
}

\NewDocumentCommand{\page}{s m}{%
  \IfBooleanTF{#1}{%
    page~\pageref*{#2}%
  }{%
    page~\pageref{#2}%
  }%
}

\IfFileExists{rev.tex}{\input{rev}}{}

\newcounter{referencepage}
\AtBeginDocument{%
  \let\oldthebibliography\thebibliography
  \renewcommand{\thebibliography}[1]{%
    \setcounter{referencepage}{\value{page}}%
    \oldthebibliography{#1}%
  }
}

\usepackage{background}
\usepackage{ifthen}

\backgroundsetup{contents={}}

\mainarxiv{
\newcommand{\maxmainpages}{11} %
}
{
\newcommand{\maxmainpages}{999} %
}
\newcommand{\ShowWarning}{%
  \ifthenelse{\value{page}>\maxmainpages\and \value{page}<\thereferencepage}{%
    \backgroundsetup{
      contents={%
        \begin{tikzpicture}[remember picture,overlay]
          \node[yshift=-0.5in] at (current page.north) {%
            \textcolor{red}{\Large\bfseries WARNING: Page limit is
            \maxmainpages~pages! It's now page \thepage .}%
          };
        \end{tikzpicture}%
      },
      angle=0,
      position={current page.north},
      scale=1,
      opacity=1
    }%
    \BgMaterial%
  }{}%
}

\AddEverypageHook{\ShowWarning}

\begin{document}

\title{Agentic Specification Generator for Move Programs}

\makeatletter
\ifdefined\DRAFT
\pagestyle{fancyplain}
\@ifpackageloaded{lastpage}{%
  \cfoot{Rev.~\therev \hfill \thepage\ of \pageref{LastPage} \hfill \thedate}%
}{%
  \@ifpackageloaded{totpages}{%
    \cfoot{Rev.~\therev \hfill \thepage\ of \pageref{TotPages} \hfill \thedate}%
  }{%
    \cfoot{Rev.~\therev \hfill \thepage\ of ? \hfill \thedate}%
  }%
}%
\fi
\makeatother

\author{
  Yu-Fu Fu\hspace{1ex}
  Meng Xu$^\dagger$\hspace{1ex}
  Taesoo Kim
  \\
  Georgia Institute of Technology \hspace{1ex}$^\dagger$University of Waterloo
}

\date{}

\sloppy

\maketitle
\arxivonly{%
  \footnotearxiv{Extended version of the same paper published on ASE'25. Extra
    appendices are added at the end of the paper.}
}
\begin{abstract}
    While LLM-based specification generation is gaining traction,
    existing tools primarily focus on mainstream programming languages
    like C, Java, and even Solidity,
    leaving emerging and yet verification-oriented languages like Move
    underexplored. 
    In this paper,
    we introduce \sys,
    an automated specification generation tool designed for Move smart
    contracts.
    \sys aims to highlight key insights that uniquely present
    when applying LLM-based specification generation to a new ecosystem.
    Specifically, \sys demonstrates that
    LLMs exhibit robust code comprehension and generation capabilities
    even for non-mainstream languages.
    \sys successfully generates verifiable specifications for 84\% of tested
    Move functions and even identifies clauses previously overlooked by experts.
    Additionally,
    \sys shows that explicitly leveraging specification language features
    through an agentic, modular design improves specification quality substantially
    (generating 57\% more verifiable clauses than conventional designs).
    Incorporating feedback from the verification toolchain
    further enhances the effectiveness of \sys,
    leading to a 30\% increase in generated verifiable specifications.
\end{abstract}

\begin{IEEEkeywords}
  LLM, Specification, Verification, Move
\end{IEEEkeywords}
\section{Introduction}
\label{s:intro}
The need for high-assurance software has been increasingly recognized
as software becomes more complex and critical.
This is especially true for blockchains and smart contracts
that now manage crypto assets worth billions of USD as
any bug or vulnerability in the code
can lead to significant financial losses for its stakeholders
\cite{defipulse2022,cryptosec2022}.
One of the key techniques to deliver high-assurance software is
formal methods such as
model checking \cite{clarke1999model} and
theorem proving \cite{bertot2004interactive}.
Among these techniques,
writing \emph{formal program specification} that captures the
intended behavior of a program is a crucial step,
as it serves as the foundation for subsequent verification tasks.

However,
writing specification is challenging,
because it requires
not only a deep understanding of the program's intended behavior
but also expressing the intention in formal languages,
which are often declarative in nature and require a different mindset
than implementing an algorithm imperatively
\cite{felderer2018formal,ferrari2022formal,
gleirscher2020formal,kulik2022survey}.
Therefore,
a more common practice in the smart contract community is that
developers often implement based on loosely documented requirements first and
then formally specify and verify the code later
(if timing and budget permit)
\cite{certora-report1, certora-report2}.
While less optimal than a waterfall model (specifications before code),
this practice is often necessary due to
limited resources and time-to-market constraints.

And yet, late specification is better than no specification at all.
To ease the burden of developers,
a practical tool that
automatically generates specifications from existing smart contract code,
with acceptable quality, would be beneficial.

In this work,
we join forces with the recent advances in
large language models (LLMs) for formal methods (LLM4FM)
\cite{autospec, specgen, xie2025effective, osvbench, dafny-loop-gen,
liu2024enhancing, kamath2023finding, chakraborty2023ranking,
liu2024propertygpt, sun2024gptscan, houdini, daikon, autoverus, SAFE,
lahirie2024evaluating, kamath2024leveraging, shrivastava2023repository}
and provide our own findings and insights
based on our experience in developing
an LLM-based tool, \sys, that
automatically generates specifications
(more precisely, functional pre/post conditions)
for smart contracts written in an emerging language---Move~\cite{move}.

\sys is based on the same rationale
why LLMs should be used for specification generation---%
manually defined specification synthesis templates~\cite{houdini,daikon}
cannot match with the diversity and complexity of programs
while LLMs can---%
as piloted in the literature
\cite{autospec,specgen,xie2025effective}.
However,
existing works fail to provide more insights on
the generalizability of their designs
nor shed light on how to adapt
to new programming/specification languages.
So in this paper,
we first seek to answer the following research questions (RQs)
through the lens of Move:
\begin{itemize}[leftmargin=8mm]
  \item[\textbf{RQ1}]
  Will LLMs show degraded
  code comprehension and generation performance
  when the programming/specification language
  is not a mainstream one
  (hence ruling out fine-tuning opportunities as well)?

  \item[\textbf{RQ2}]
  As specifications need to be expressed in a formal language,
  what features of the specification language
  LLMs can leverage to generate better specifications?

  \item [\textbf{RQ3}]
  As the generated specifications will be eventually
  verified by a verification tool to prove that
  the code satisfies the specifications,
  will the feedback of such a verification tool help
  improve the quality of generated specifications?
\end{itemize}

Move is a suitable candidate for these RQs
because it is a relatively new language
(debuted in 2020 for the discontinued Diem blockchain~\cite{diem}
and later evolved by Aptos~\cite{aptos}, Sui~\cite{sui}, and
Movement~\cite{movement} blockchains).
Compared with mainstream ones like C, Java, or even Solidity
which are preferred in prior
works~\cite{autospec,specgen,liu2024propertygpt,osvbench,xie2025effective},
Move has very limited codebases or documentation for LLMs to train or fine-tune
on.
And yet,
formal verification is a first-class citizen in Move.
In fact, Move comes with a built-in specification language
called Move Specification Language (MSL),
which allows developers to write expressive specifications for their code.
The specifications are checked by an automatic verification tool---%
Move Prover~\cite{move-prover}---%
which discharges proof obligations to
SMT~\cite{smt} solvers like Z3~\cite{z3} or CVC5~\cite{cvc5}.

However,
while addressing RQ1--3
help establish the feasibility of \sys,
in order for \sys to be practical and useful
in real-world Move codebases,
we still need to resolve two technical concerns
that are unaddressed in prior works:
\begin{itemize}[leftmargin=8mm]
  \item[\textbf{RQ4}]
  How to properly scope each LLM conversation context
  if the ultimate goal is to
  autonomously generate a useful set of specifications
  for the entire codebase without interrupting developers
  (i.e., human-in-the-loop)?

  \item[\textbf{RQ5}]
  While the accuracy (or quality) of
  generated specifications can be evaluated by
  comparing them against manually written specifications,
  how can we evaluate the comprehensiveness
  of the generated specifications?
\end{itemize}

\PPI{Key Findings.}
The design of \sys is heavily influenced by the answers to these RQs,
which we summarize as follows:

\begin{itemizesqn}
\item \uline{LLMs show remarkable performance in
  comprehending Move code and generating Move specification (in MSL syntax)
  despite the language's recency (\textbf{RQ1}).}
  Overall,
  \sys successfully generates specifications for
  84\% (300/357) Move functions in Aptos core
  (analogous to \cc{libc} in C) with the OpenAI o3-mini model.
  Compared to expert-written specifications, \sys generates 82\% of matching
  verification conditions and produces an additional 57\% that differ;
  aggregately
  39\% more than the expert-written ones.
  This is on-par with results in prior works on different languages:
  \textit{SpecGen} \cite{specgen} shows overall 60\% correctness on Java, while
  \textsc{AutoSpec} \cite{autospec} achieves 79\% correctness on C.

\item \uline{Specification language matters and leveraging its features
  explicitly at tool design time (instead of implicitly by LLMs)
  can improve the quality of generated specifications (\textbf{RQ2}).}
  MSL allows function pre/post conditions to be expressed in four
  different classes of specifications,
  each encodes developers intentions from a different angle.
  \sys actively leverages this by
  generating sub-specifications of different classes in MSL,
  with one specialized agent for each class,
  and subsequently \emph{ensemble} sub-specifications
  into an idiomatic one.
  Our evaluation shows that the
  agentic design generates 57\% more verifiable specifications
  than a conventional all-in-one design,
  which implicitly relies on LLMs to learn the differences
  between the classes of specifications.

\item \uline{Feedback from the verification tool
  can significantly improve the quality of generated specifications
  (\textbf{RQ3}).}
  While prior works have shown that
  leveraging compiler feedback help generate
  syntactically correct specifications~\cite{liu2024propertygpt},
  \sys goes a step further by
  leveraging the feedback from Move Prover (especially the counterexample)
  as an oracle to fix the wrongly generated
  specifications (similar to~\cite{specgen, kamath2024leveraging}),
  which improves the accuracy of
  generation as shown in the evaluation results.
  We observe 30\% more verifiable specifications when using the Move Prover
  feedback for all-in-one design in the ablation study of our evaluation.
  When prover feedback is removed from the agentic design, the accuracy does not
  dramatically drop because of the fail-safe design: 13.1\% less function are
  verified.
  However, a clear decrease is observed in the quality of generated
  specifications: it generates 33\% less of clauses than the agentic design with
  prover feedback.

\item \uline{Scaffolding is necessary to
  keep LLMs focused on the context of the function being specified
  while also catering to the compositional nature of
  functional specification (\textbf{RQ4}).}
  Sending the entire codebase to LLMs is obviously
  infeasible~\cite{shrivastava2023repository},
  and sending only the function being specified is often not enough,
  \sys implements a series of scaffolding utilities including
  static dependency analysis,
  selective function inliner, and
  a pretty-printer to lift Move abstract syntax tree (AST)
  back to source code after inlining.
  These utilities help \sys to scope the conversational context
  with LLMs autonomously without human-in-the-loop.
  For example, \sys can automatically decide whether inline the
  callee functions based on the best scoping strategy.
  During the evaluation, selective inlining unblocks 7 new
  functions for agentic design and 20 new functions for all-in-one design,
  respectively.

\item \uline{Specification coverage,
  a simple metric to measure
  how much code in the function body is covered by specifications,
  can be used to evaluate comprehensiveness (\textbf{RQ5}).}
  Our experiment shows that \sys generated unique clauses that were missed in
  expert-written specifications: 130 \cc{ensures}, 86 \cc{aborts\_if}, and 75
  \cc{modifies} clauses.
  These account for 33.2\% (291) of the total 876 clauses generated by \sys.
  Furthermore, the specification coverage component validates its outstanding
  performance, as \sys produces comprehensive specifications covering provided
  code.
\end{itemizesqn}

\section{Background}
In this section, we briefly describe the
Move programming language,
Move specification language (MSL), and
Move Prover.

\label{s:background}
\subsection{Example for Move Smart Contract and Move Specification}
\begin{figure*}[ht]
  \centering
  \begin{subfigure}[t]{0.48\linewidth}
    \input{code/transfer.move}
    \caption{Move Smart contract for a Coin}
    \label{fig:transfer}
  \end{subfigure}
  \hfill
  \begin{subfigure}[t]{0.48\linewidth}
    \input{code/transfer.spec.move}
    \caption{Corresponding specification}
    \label{fig:transfer-spec}
  \end{subfigure}
  \caption{An example of Move smart contract and its
  corresponding specification for a simple Coin used in Move-based blockchains.}
  \label{fig:move-example}
\end{figure*}

To get a taste of programming and specifying in Move,
\autorefp{fig:move-example} shows a simple Move smart contract and
its specification.
In the implementation (\autoref{fig:transfer}),
two data types, \cc{Coin} and \cc{Balance},
and three functions, \cc{transfer}, \cc{withdraw}, and \cc{deposit},
are defined.
\cc{transfer} withdraws a coin from the sender's account and deposits it into
the receiver's account by internally calling \cc{withdraw} and
\cc{deposit} respectively.
In \cc{withdraw} and \cc{deposit}, \cc{borrow\_global\_mut} is used to borrow a
mutable reference to the global state of the blockchain, which is a map from
addresses to the data structure \cc{Balance}.

The corresponding specification (\autoref{fig:transfer-spec})
defines the intended behavior of these functions,
which follows the standard style of Hoare triple \cite{hoare1969axiomatic}
with precondition (\cc{true}) omitted
(although more complicated pre-conditions
can be specified with \cc{require} clauses).
\cc{aborts\_if} clauses capture
exhaustively the conditions
under which the function should abort.
The \cc{modifies} clauses marks modifications to
blockchain global state,
and the effects of such modifications are often captured by
\cc{ensures} clauses in an axiomatic style.
Collectively,
\cc{modifies}, \cc{aborts\_if}, and \cc{ensures} clauses
capture the complete set of post-conditions for the specified function.

As a concrete example,
the specification of \cc{transfer} consists of two parts:
bindings (by \cc{let} and \cc{let post}) and properties.
For bindings (Line 3-9),
\cc{global} is used to access the global
state of the blockchain,
which is an immutable (read-only) reference.
\cc{let} is used to bind the state of variables \emph{before} the function is
executed, and \cc{let post} is used to bind the state of variables \emph{after}
the function is executed.
The \cc{transfer} function modifies the global states
of the struct \cc{Balance} because of its calls to \cc{withdraw} and
\cc{deposit}, which both also modify the global state of \cc{Balance}
(Line 10).
The \cc{transfer} function should abort when
the \cc{Balance} struct does not exist for either sender or receiver,
on overflow,
or when the sender has insufficient balance (Line 12-16).
The \cc{transfer} function, upon successful execution,
should ensure that the sender's balance is reduced by the amount transferred and
the receiver's balance is increased by the same amount (Line 18-19).

\subsection{Move Prover}
Move Prover \cite{move-prover} is an \emph{automatic} formal verification tool
specifically designed for Move,
which as of now is primarily used within the Aptos blockchain ecosystem
to ensure the correctness and security of smart contracts.
Move Prover first translates specification and implementation
into Boogie intermediate verification language~\cite{boogie},
based on which the actual verification conditions (VCs) are produced
in the form of SMT formulas,
which are subsequently checked by SMT solvers
such as Z3~\cite{z3} or CVC5~\cite{cvc5}.
One notable features of Move Prover is
its ability to provide feedback in
the form of a call stack in Move source code
when encountering verification failures
by lifting the counterexample provided by Boogie
back to the Move source code level.
The original intention of this feature is to help developers
debugging issues in code or specifications but
such feedback can also be analyzed by LLM to auto-fix generated specifications,
as later shown in \autoref{s:eval}.

\section{\sys Design}
\label{s:design}
\newcommand{\contf}{$\textit{context}_{f}$\xspace}
\newcommand{\calleef}{$\textit{callee}_{f}$\xspace}
\newcommand{\contcalleef}{$\textit{context}_{\textit{callee}_{f}}$\xspace}
\newcommand{\fun}{$f$\xspace}
\newcommand{\spec}{$s$\xspace}
\newcommand{\ce}{$c$\xspace}
\newcommand{\cov}[1]{\textit{cov}_{#1}}

\begin{figure*}[ht]
  \centering
  \includegraphics[width=0.8\linewidth]{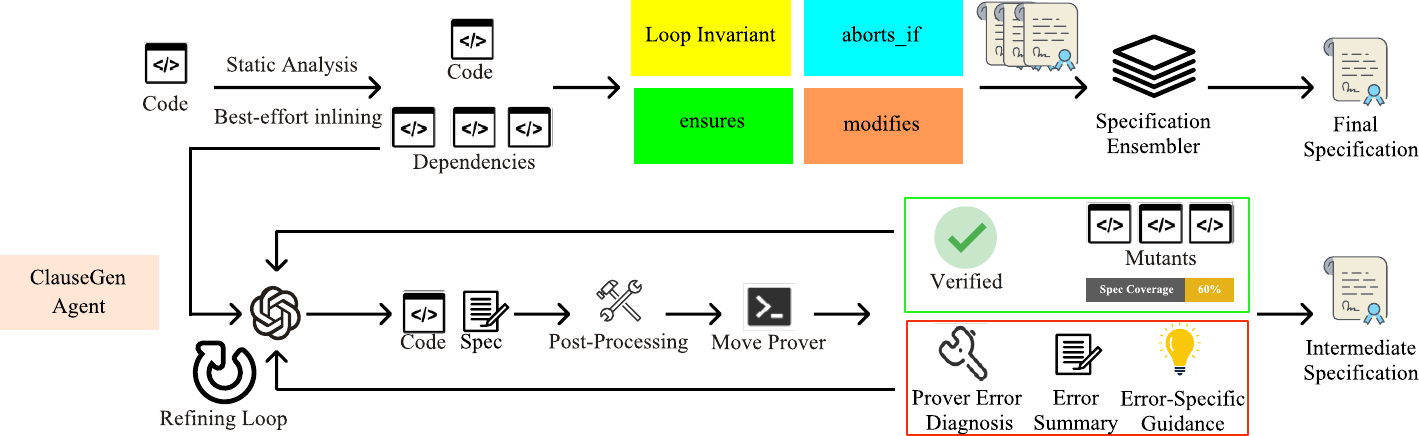}
  \caption{Workflow of \sys. The top row shows the main agentic
    workflow of \sys, which contains 4 clauses generation (ClauseGen)
    agents for each class of specifications and the specification ensembler for
    idiomatic Move specifications.
    The bottom row shows a workflow in a single ClauseGen
    agent, which runs a generation loop to refine or fix the generated
  specification.}
  \label{fig:overview-agent}
\end{figure*}

\autorefp{fig:overview-agent} illustrates the overall agentic design of \sys.
To kickstart \sys,
users specify the target Move function
they want to generate specification for.
\sys then performs static analysis
to extract the function and its dependencies,
such as definitions to
(direct and indirect) callees,
data structures, and
constants involved.
This enables \sys to produce two versions of conversation context
for subsequent LLM interactions:
(\textbf{V1})
target function with all callees inlined at best-effort
(\autoref{ss:context})
which often leads to a single function with a complicated function body;
(\textbf{V2})
target function with all callees' definition listed alongside,
leveraging LLM's inherent capability to recognize function calls.
Both versions are dispatched to separate
specification generation (ClauseGen) agents (\autoref{ss:design-agents})
as seed conversation context.

\sys employs four different ClauseGen agents,
each specialized in generating a specific class of
functional specification clauses expressible in MSL:
loop invariants (if the code contains loops),
abort conditions (i.e., \cc{aborts\_if} clauses),
global state modification markings (i.e., \cc{modifies} clauses), and
axiomatic semantics for state changes (i.e., \cc{ensures} clauses).
Generated specification snippets from all agents
are subsequently merged and optimized by
a specification ensembler
to produce a coherent and idiomatic set of specifications.
While an alternative design
could generate all classes of clauses in a single pass,
\sys adopts an agentic design to avoid overwhelming LLM with too many
instructions in a single system prompt,
which is later proven effective in our evaluation (\autoref{ss:ablation-study}).

Inside each ClauseGen agent,
\sys runs a generation loop to refine or fix the
generated specification.
After each round,
the generated clauses are first vetted by the
post-processing steps to fix common issues
(e.g., misplaced semicolon or spaces), if any,
followed by Move Prover to check
whether the vetted specification, after ensembling,
pass verification or fail with a diagnostic message
(e.g. syntax error or a counterexample) (\autoref{ss:prover-feedback}).
If the ensembled specification fails,
\sys will first instruct the LLM to summarize
the diagnostic message and generate guidance for possible fixes.
Besides the LLM-generated guidance,
\sys also includes pre-defined guidance for
common error patterns.
All this information will be included in the user prompt for the next round of
generation loop to help corresponding ClauseGen agent(s)
fix wrong clauses.
On the other hand,
if the generated specification passes Move Prover,
\sys will generate a series of mutated code snippets
by randomly deleting parts of original Move code
(i.e., nodes in the AST)
to get the specification coverage (\autoref{ss:spec-cov}),
which will be included in the user prompt for the next round
to refine the specification.
The whole process is repeated
until a predefined bound (e.g. 5 rounds).

\PN{``Agentic'' Terminology.}
\phantomsection\label{rev-agentic}
We treat self-contained LLM components as distinct sub-agents—such as
ClauseGen (which includes clause generation and error summary/guidance) and
the specification ensembler—each handling specific sub-tasks.
While the definition of ``agent'' is debated in communities, our approach uses
these sub-agents in a controlled, orchestrated manner rather than through a
dedicated autonomous planning agent.
Given \msg's modular architecture, integrating a planning agent would be
straightforward.
Notably, ClauseGen adopts an agent pattern—the Reflection
pattern~\cite{agent-reflexion}—which enables it to fix errors autonomously
with the help of Move Prover diagnostics and LLM-powered error summaries that
suggest possible fixes.

\subsection{Agentic Design for Different Classes of Specifications}
\label{ss:design-agents}

As previously explained,
\sys uses specialized agents to
generate four classes of clauses expressible in MSL---%
\cc{aborts\_if},
\cc{modifies},
\cc{ensures},
and loop invariants---%
independently.
This design decision is motivated by the observation that
generating all types of clauses in one-pass could be too burdensome for LLM.
To be specific, in an all-in-one design,
we need to pack comprehensive MSL guidelines in a \emph{single} system prompt,
in which not all instructions are closely followed by the LLM.
To make things worse,
failures to generate one class of clauses might lead
to overall failures in the all-in-one design,
which is not ideal for a specification generation tool---ideally,
even if one class of clauses fails,
the rest of the clauses might still be correct and useful for users.

Backed by the fact that different classes of clauses encode
completely different aspects of the function semantics
(i.e., they are inherently compositional) in MSL,
\sys takes the stance that
there is no need to generate them in a single pass.
Instead,
\sys devises an agentic design:
specific ClauseGen agents with specialized prompts for each class of
clauses:

\PN{\cc{aborts\_if} Clauses.} This agent is to find possible scenarios of
aborts, such as
integer overflows (e.g., line 15 in~\autoref{fig:transfer-spec}),
missing resources (e.g., line 12 in~\autoref{fig:transfer-spec}),
or manually annotated abort conditions in the form of assertions.

We instruct LLM to carefully looks
for various types of abort conditions to generate \cc{aborts\_if} clauses.
Additionally, we independently run Move Prover to check
whether \cc{aborts\_if false} is true.
\cc{aborts\_if false} signals the intention that
the function will never abort in any condition,
which is the strongest predicate for this class of clause.
Therefore,
if the Move Prover passes the \cc{aborts\_if false} check,
we skip abort condition generation and simply output \cc{aborts\_if false}.

\PN{\cc{modifies} Clauses.}
When \cc{borrow\_global\_mut} expressions are identified
in the target function (and its callees),
\sys launches this agent to find all such expressions,
check whether the mutable references are actually written, and
generate corresponding \cc{modifies} clauses.
In Move, if we need to verify that the caller function modifies the global
state, we need to have specifications for the callee functions, too.
For example, in \autoref{fig:transfer-spec}, the function \cc{transfer} modifies
the \cc{Balance} global resources because of the callees \cc{withdraw} and
\cc{deposit}.
To pass the Move Prover, in addition to the \cc{modifies} (Line 10) in
\cc{transfer}, the ones in \cc{withdraw} (Line 23) and \cc{deposit} (Line 35)
are also needed.
Therefore, in the specialized system prompt for \cc{modifies} clauses, we
instruct LLM to also generate the \cc{modifies} clauses for the callee
functions.

\PN{\cc{ensures} Clauses.} This agent is used to generate the post-conditions of
the function: for example, a function summary which captures the correspondence
between the function inputs and outputs like the \cc{transfer} example in
\autoref{fig:transfer-spec}.
We instruct the LLM to identify operations linking input (function arguments),
output (return values), and global state modifications.
When a function modifies the global state, it should use \cc{let} and \cc{let
post} to capture the state before and after execution.
We also provide a prompt addressing common syntactic errors specifically in
\cc{ensures} clauses.
For example, if a function contains branches (if-else), the LLM might generate
disallowed constructs such as \cc{ensures if (c) ... else ...}.
Instead, we suggest encoding branches with the \cc{==>} (implies) operator:
\cc{ensures c ==> ...; ensures !c ==> ...}, which significantly reduces errors
when dealing with branches.

\PN{Loop Invariants.} When the code context contains loops,
we enable this agent to find loop invariants.
However,
unlike other agents that simply return the requested clauses,
this agent will return an
\emph{annotated} function with loop invariants embedded in the function body
(as required by the MSL syntax).
In addition, loop invariant inference by itself is a complex task and an
independent research topic even in LLM4FM~\cite{dafny-loop-gen,liu2024enhancing,
  kamath2023finding, chakraborty2023ranking}.
To generate loop invariants, we instruct the LLM to carefully inspect the loop
body, identifying modified variables and preserved properties as candidates for
invariants.
We observed many syntactic errors in the generated invariants, largely due to
limited training data for Move—especially for loop invariants.
Therefore, in addition to an annotated function example with loop invariants, we
include a set of invariant examples to guide the LLM.

\PP{Ensembler for Idiomatic Specification}
Although MSL permits embedding all pre/post conditions and loop
invariants in the code like other deductive verification tools
with inline specification blocks, it is not the idiomatic way in Move.
\emph{Idiomatic} Move specification instead puts
all clauses in a separate \cc{spec} block (or even file)
except for loop invariants.
Additionally, in the \cc{spec} block,
MSL supports declaring variable bindings
(\cc{let}/\cc{let post}) for convenience and clarity.
Therefore,
when we have multiple specification from the ClauseGen agents discussed
above, although they specify different properties of the function,
they might declare overlapping variable bindings,
which is not allowed.
Naively concatenating bindings and clause snippets from different
ClauseGen agents leads to not only compilation errors but also an
over-verbose set of specification.
As a result, \sys uses an ensembler (another LLM agent)
to merge the output from ClauseGen agents.
On top of that, the ensembler also enforces
coherent coding styles,
such as ordering
variable bindings,
\cc{modifies}, \cc{aborts_if}, and finally \cc{ensures} clauses
in the \cc{spec} block.

\subsection{Verifier-in-the-Loop: Incorporating Move Prover Feedback}
\label{ss:prover-feedback}

While powerful,
LLMs can rarely generate an acceptable set of specification in one attempt.
Failures may arise due to compilation errors (e.g., wrong syntax) or
semantic errors (incorrect specifications).
As LLMs cannot self-validate the generated specifications,
we use Move Prover as an oracle for verification.
Move Prover checks for compilation errors and, if none exists,
attempts to prove that the code matches the generated specification.
This process will yield one of the following outcomes:
\begin{enumerate*}[label=\protect\BC{\arabic*}]
\item the generated specification passes the prover,
\item the generated specification fails with a counterexample generated,
\item the verification times out, or
\item a compilation error is reported.
\end{enumerate*}
As long as verification fails (\BC{2},\BC{3},\BC{4}),
we include the
Move Prover diagnostics,
error summary,
pre-defined guidance on common fixes for known error types, and
the counterexample (if any)
in the prompt for one or more responsible ClauseGen agents
to re-generate a new set of clauses in the next round.

More specifically,
if the verification failure is caused by an incorrect \cc{ensures} clause,
\sys will instruct the \cc{ensures} agent
(and the loop invariant agent if loops are involved)
to re-generate a new set of clauses.
\sys also attempts to provide some pre-defined
guidances for common errors by pattern matching the prover error message.
The most common errors include:
the usage of undefined functions or impure functions
(e.g. containing early returns in CFG)
that cannot be used in the specification because they cannot be translated to
Boogie.
We use regular expressions to match the prover error message to find
those bad function usages and guide LLM to avoid them in the next round of
generation loops.

\subsection{Specification Coverage to Pinpoint Missing Parts}
\label{ss:spec-cov}
\begin{figure*}[ht]
  \centering
  \begin{subfigure}[t]{0.48\linewidth}
    \input{code/random-deletion.move}
    \caption{Original Move Code $f$}
    \label{fig:original-move}
  \end{subfigure}
  \hfill
  \begin{subfigure}[t]{0.48\linewidth}
    \input{code/random-deletion-mutate.move}
    \caption{Modified Move Code $f^{-}$}
    \label{fig:modified-move}
  \end{subfigure}

  \vspace{1em}

  \begin{subfigure}[t]{0.48\linewidth}
    \input{code/random-deletion-spec-complete.move}
    \caption{Complete Specification $s^{+}$}
    \label{fig:complete-spec}
  \end{subfigure}
  \hfill
  \begin{subfigure}[t]{0.48\linewidth}
    \input{code/random-deletion-spec-wrong.move}
    \caption{Incomplete Specification $s^{-}$}
    \label{fig:incomplete-spec}
  \end{subfigure}

  \caption{Example to show the random deletion of Move code. (a) Original Move
    code, (b) Modified Move code, where the deleted parts are marked with a
  comment, (c) Complete specification, and (d) Incomplete specification.}
  \label{fig:example-random-deletion}
\end{figure*}

Generating a complete (strong) specification for a function is notoriously hard.
Therefore,
Move Prover accepts partial specifications and
expressions in MSL have partial semantics by design.
This implies that there might be multiple specifications
that can pass the verification for a given function.
To differentiate and rank them,
we introduce a metric called \emph{specification coverage},
analogous to traditional line code coverage.
Intuitively,
given two specifications, $s_a$ is considered more complete than $s_b$
if $\cov{s_a} > \cov{s_b}$.

Unlike code coverage which can be readily collected during execution,
we cannot instrument specifications and track how they are used in the proof
obligations. 
Instead,
we approximate specification coverage via a method called
\emph{random deletion} (inspired by \textsc{Fast}~\cite{fast}).
The idea is simple: if the implementation is incorrect after deleting parts of
code, then a correct specification $s$ should fail the Move Prover, unless $s$
is incomplete and captures only part of the behavior.

Consider the function $f$ in \autorefp{fig:original-move} with its
complete specification $s^{+}$ in \autoref{fig:complete-spec} (verifying both
elements of a pair), versus a modified function $f^{-}$ in
\autoref{fig:modified-move} and its incomplete specification $s^{-}$ in
\autoref{fig:incomplete-spec} (verifying only one element). Here, $s^{-}$ can
pass the prover for both $f$ and $f^{-}$, while $s^{+}$ passes only for $f$,
indicating that $s^{+}$ is more complete.

We implement random deletion at the AST level,
and the specification coverage measurement is as follows:
\begin{enumeratesqn}[label=\protect\WC{\arabic*}]
\item Randomly delete parts of the AST (e.g., code blocks, statements,
  expressions) of function to create mutants.
\item Lift modified AST to source code and
  check the code mutant against the generated specification by Move Prover.
\item If verification passes, the deletion is not covered by the
  specification; otherwise, the part is covered.
\end{enumeratesqn}
We then include line diffs for the deleted codes to pinpoint the non-specified
parts in the prompt for the next round.

\subsection{Construct Context for Specification Generation}
\label{ss:context}
The full dependency information for the target function is indispensable for
verifiable specification generation.
While including the entire Move library is tempting, this is
unrealistic given LLM context length limits and instruction-following
capabilities.
To address this, \sys uses static analysis to slice the target and its
dependencies from the library and packs them as the normal conversion context
(\textbf{V2}).
However, even after slicing, the isolated dependencies might still be
overwhelming.
Thus, we perform best-effort function inlining to selectively fold as many
dependencies as possible into the target function’s body, reducing LLM burden in
reasoning across multiple function boundaries.
The inlined target function and remaining non-selected dependencies are then
packed as the inlined conversion context (\textbf{V1}).
Both contexts are dispatched to ClauseGen agents for specification generation.

\PP{Static Analysis for Cross-module Function Dependencies}
We perform interprocedural def-use analysis to identify target functions and
their dependencies (structs, constants, and both direct and indirect callees).
Using this analysis, \sys tracks \emph{cross-module usage} and uncovers
dependencies often missed by similar tools limited to single code snippets.

\PP{Best-effort Function Inlining to Reduce Context Complexity}
\sys intercepts compilation to eagerly inline callees.
Naively inlining all callees is infeasible due to compiler safety checks (e.g.,
resource safety).
Instead of exhaustively enumerating safe combinations, \sys uses a
\emph{best-effort} monotonic approach: it attempts to inline the first callee,
inlining it if compilation passes; if not, it skips it.
This continues for each callee, with later ones skipped if they conflict with
earlier inlines.

The inlined representation is then generated as an AST, converted back to Move
code via an AST pretty-printer, and used to build the inlined conversion context
(\textbf{V1}).

\subsection{Abstract Specification as Last Resort}
\label{ss:abstract-spec}
When all ClauseGen agents fail to generate a specification,
e.g., due to excessive complexity or unsupported operations
(e.g., non-linear arithmetic),
MSL allows writing abstract specifications~\cite{aptos:abstract-spec}
using \emph{uninterpreted functions}~\cite{uninterpreted-function}
as stubs.
These stubs may later be replaced with concrete specifications or serve as
implementation contracts.
Thus, instead of abandoning the effort when the LLM fails across all
specification classes, \sys instructs the LLM to generate abstract
specifications as proof templates to assist experts.
However, we do not consider abstract specifications as verifiable specifications
in evaluating \sys as they are only placeholders for developers essentially.
\arxivonly{%
  \autoref{app:abstract-spec-example} shows a concrete example of an abstract
  specification generated by \sys.}

\section{Implementation}
\label{s:impl}

\sys is implemented within the
Aptos Move toolchain monorepo
for ease of integration and code reuse.
\sys consists of the core agentic system
that generates, fixes, and refines specifications (5k LoC)
as well as scaffolding utilities
(e.g., dependency analysis, pretty-printer, scripts) consisting of 2.3k LoC.
The entire monorepo with \sys included can be found
in the supplementary material associated with this paper.
\sys is open source at \url{https://github.com/sslab-gatech/MSG}.

\section{Evaluation}
\label{s:eval}

We evaluate \sys based on the research questions (RQs)
that both motivate and influence its design (\autoref{s:intro}):

\begin{itemizesqn}
\item \textbf{RQ1:}
(\emph{Effectiveness})
How effective is \sys in
generating acceptable specifications for real-world codebases?
(\autoref{ss:eval-result})

\item \textbf{RQ2, RQ3, RQ4:}
(\emph{Ablation Study}):
  How does each component affect the performance of \sys
  (\autoref{ss:ablation-study})?
  More specifically:
  is the agentic design of \sys beneficial?
  is the Move Prover feedback useful?
  is function inlining helpful?
\item \textbf{RQ5}
(\emph{Comprehensiveness}):
  How complete are the specifications generated by \sys?
  Are there gaps in coverage (\autoref{ss:eval-cov})?
\end{itemizesqn}

\PP{Experiment Settings}
\sys attempts to generate specifications in multiple rounds
as shown in~\autoref{fig:overview-agent}.
During this evaluation, we set the number of rounds to 5.
We conduct the same experiments 3 times to reduce randomness from the LLM.

\PP{Target Move Codebase}

We have evaluated \sys on the following
foundational Move libraries on Aptos:
\begin{itemizesqn}
\item move-stdlib: standard libraries like vector, string.
\item aptos-stdlib: extended standard libraries designed for Aptos.
\item aptos-framework: Aptos Framework libraries including account, coin, etc,
  which are used to build the smart contracts.
\end{itemizesqn}
\noindent
Sui~\cite{sui}, another popular Move-based blockchain mentioned in
  \autoref{s:intro}, is not chosen as the Move Prover is not compatible with the
  extended UTXO model embedded in Sui Move.

\PP{Target Function Selection}
Not all the functions in the target Move codebase are selected
as the target for specification generation.
We exclude
\begin{enumerate*}
\item native functions which are implemented in Rust instead of Move.
\item functions that are marked as not verified (marked with \cc{pragma verify = false}).
\end{enumerate*}
\autoref{tab:benchmark} presents the breakdown details for each library:
  357 functions in total, 17 of them containing loops.
The small number of loop-related functions arises because: (1) smart
  contracts seldom use loops, and (2) loops mostly occur in low-level libraries,
  such as vector modules.

\begin{table}[ht]
  \centering
  \begin{tabular}{lccccc}
    \toprule
    \textbf{Codebase} & \textbf{LOC} & \textbf{Public} & \textbf{Private}
    & \textbf{Args} & \textbf{Loops} \\
    \midrule
    \textbf{move-stdlib}  & 82    & 10  & 0  & 15  & 8  \\
    \textbf{aptos-stdlib}       & 397   & 58  & 5  & 122 & 1  \\
    \textbf{aptos-framework}    & \num{2780} & 205 & 79 & 492 & 8  \\
    \midrule
    \midrule
    \textbf{TOTAL}              & \num{3259} & 273 & 84 & 629 & 17 \\
    \bottomrule
  \end{tabular}
  \caption{Code metrics for three Move codebases.
      357 functions in total,
      17 of them containing loops.
      LOC counts only lines inside function bodies
      (type/constant declarations excluded).
      Public/Private are function counts;
      Args is the total number of formal parameters across all functions; Loops
      counts functions with loops.
    }
  \label{tab:benchmark}
\end{table}

\PP{Specification Quality Metrics}
We use two metrics to gauge the quality of generated specifications:
verifiability and comprehensiveness.
Verifiability simply means that
the generated specification,
combined with the implementation of the target function,
produces valid verification conditions that can be proved by the Move Prover.

Regarding specification comprehensiveness,
Aptos smart contracts
include expert-written Move specifications for parts of their
codebase.
To assess whether the specifications generated by \sys are comparable to these
experts' ground-truth specifications, we define a metric called
\emph{specification comprehensiveness}, which is checked by
  subsumption~\cite{subsumption}.

Consider the complete specification $s^+$ in
\autoref{fig:complete-spec}, which proves both the first and second
members of the result pair, versus an incomplete specification
$s^{-}$ in \autoref{fig:incomplete-spec}, which only proves the first
member. In this case, the complete specification is stronger, as it
implies the incomplete one ($s^+ \implies s^{-}$).

We formalize this idea as follows. Given a generated specification $s_g$ and an
expert-written specification $s_e$, we first decompose $s_e$ into
verification conditions:
\[
  s_e \equiv s_{e_1} \land s_{e_2} \land \cdots \land s_{e_n},
\]
where each $s_{e_i}$ corresponds to clauses such as \texttt{ensures}
and \texttt{aborts\_if} in Move. We then check for each $i \in
\{1,2,\ldots,n\}$ whether $s_g$ implies $s_{e_i}$ and count the
number of successful implications. The ratio of successful cases over
$n$ serves as the specification comprehensiveness.

Automatic splitting and implication checking is non-trivial.
Fortunately, with a manageable number of functions (357), we manually
extracted and split the expert-written specifications from the Aptos
Move codebase, and verified each implication through human inspection
or by running the Move Prover.

\PP{LLM Backends}
During the evaluation, we use OpenAI's LLMs as the backend for \sys.
We use the official API endpoints without web-browsing features.
Both base models and reasoning models are used, as follows:
\begin{itemizesqn}
\item OpenAI o3-mini: the compact reasoning model with default medium reasoning
  effort (o3-mini-medium).
\item OpenAI GPT-4o: the general-purpose model.
\item OpenAI GPT-4o-mini: the compact general-purpose model.
\end{itemizesqn}

\noindent During our evaluation, the weaker models, OpenAI GPT-4o and
GPT-4o-mini, are only used by the simpler design, \sysaio, for the
ablation study (\autoref{ss:ablation-study}) to showcase that stronger
  models have better performance in generation as expected.
  \mainarxiv{%
    Due to space limitations, evaluation results for weaker models (GPT-4o and
    GPT-4o-mini) are provided in the extended version~\cite{msg-extended}. Only
    results for o3-mini are included in this version.}%
  {We also evaluate weaker models (GPT-4o and GPT-4o-mini) in addition to
    o3-mini results (see \autoref{tab:ablation} and \autoref{tab:inline}).}

\begin{table*}[hbt]
  \centering
  \begin{subtable}[t]{0.45\textwidth}
    \centering
    \begin{tabular}{cccc}
      \toprule
      & \sys & \sysm & \sysinline \\
      \midrule
     Fail & 0 & 0 & 0 \\
     Success & 300 & 253 & 299 \\
     Abstract & 57 & 104 & 58\\
    \midrule
    \midrule
    \cc{ensures}     & 466  & 328   & 510 \\
    \cc{aborts\_if}    & 252  & 182   & 245  \\
    \cc{modifies}      & 144  & 73  &  89 \\
    \cc{Loop invariants}  & 14  & 7   & 14 \\
    \textbf{All clauses}      & 876  & 590     & 858 \\

      \midrule
      \midrule
      $\frac{\text{Success}}{Total}$ & \textbf{84\%} & 70.9\% & \textbf{83.8\%} \\
      \midrule
       \% of Clauses &  \textbf{100\%} & 67.6\% & 97.9\% \\
      \bottomrule
    \end{tabular}
    \label{tab:agent-results}
  \end{subtable}
  \hfill
  \begin{subtable}[t]{0.54\textwidth}
    \centering
    \begin{tabular}{c cccc}
      \toprule
      & \sysaio & \sysaiom & \sysaioinline & \sysaionaive \\
      \midrule
      \nth{1}   & 132 & 140 & 132 & 106 \\
      \nth{2}   & 33 & 6 & 35 & 18 \\
      \nth{3}   & 14 & 0 & 15 & 7 \\
      \nth{4}   & 8 & 0 & 7 & 7 \\
      \nth{5}   & 4 & 1 & 4 & 3 \\
      \midrule\midrule
      Fail         & 54 & 77 & 47 & 220 \\
      Success      & 191 & 147 & 193 & 137 \\
      $ > \nth{1}$ & 59 & 7 & 61 & 31 \\
      Abstract     & 112 & 133 & 117 & 0 \\
      \midrule\midrule
      $\frac{\text{Success}}{Total}$ & 53.5\% & 41.1\% & 54.1\% & 38.3\% \\
      $\frac{\nth{1}}{\text{Success}}$       & 69.1\% & 95.2\% & 68.4\% & 77.4\% \\
      $\frac{> \nth{1}}{\text{Success}}$       & 30.9\% & 4.8\% & 31.6\% & 22.6\% \\
      \midrule
      $\frac{\text{Success}}{\text{Success} + \text{Abstract}}$       & 63\% & 52.5\% & 62.3\% & 100\% \\
      \bottomrule
    \end{tabular}
    \label{tab:variants-detail-results}
  \end{subtable}

  \caption{
      Comparison of agentic design (\sys variants) and all-in-one design
      (\sysaio variants) using o3-mini model.
      \newline
    \textbf{Left:} Results of \sys, \sysm, and \sysinline with counts of
    failures,
    successes, and abstracts specification along with clause metrics
    (\cc{ensures}, \cc{aborts\_if}, \cc{modifies}), loop invariants, and success
    rates.
      \newline
    \textbf{Right:} Detailed results for all-in-one variants showing
    round-by-round
    performance where \nth{1} to \nth{5} represents the earliest successful
    round.
    ``Fail'' means the system cannot generate specification within five rounds.
    ``Success'' is the sum of \nth{1} to \nth{5}. ``$> \nth{1}$'' means
    non-one-shot generation (sum of \nth{2} to \nth{5}).
    ``Abstract'' is the number of generated abstract specifications.
    The success rate, one-shot rate, and concrete specification rate are shown
    in the bottom rows.
}
  \label{tab:merged-results}
\end{table*}

\subsection{Overall Effectiveness of \sys}
\label{ss:eval-result}

\PP{Specification Verifiability}
As shown in~\autorefp{tab:merged-results},
\sys
successfully generated valid specifications for
\emph{84\%} of the target functions (300 out of 357),
which is a clear indication that \sys is effective in generating specifications
for real-world Move codebases.
The specifications generated by \sys cover
all 10 functions in move-stdlib,
48 out of 63 in aptos-stdlib, and
242 out of 284 in aptos-framework as shown in
\autorefp{tab:spec-stats-o3-mini-conditions}.
Additionally,
\sys generates abstract specifications (\autoref{ss:abstract-spec})
for the remaining 57 functions,
which serve as placeholders for experts to further complete.
Finally, within the 357 functions, 17 of them contain loops. \sys
  successfully generates verifiable loop invariants for 14 out of 17
  functions, which is at the same level of accuracy as generation for loop-free
  functions.

\begin{figure}[ht]
  \centering
  \includepdf{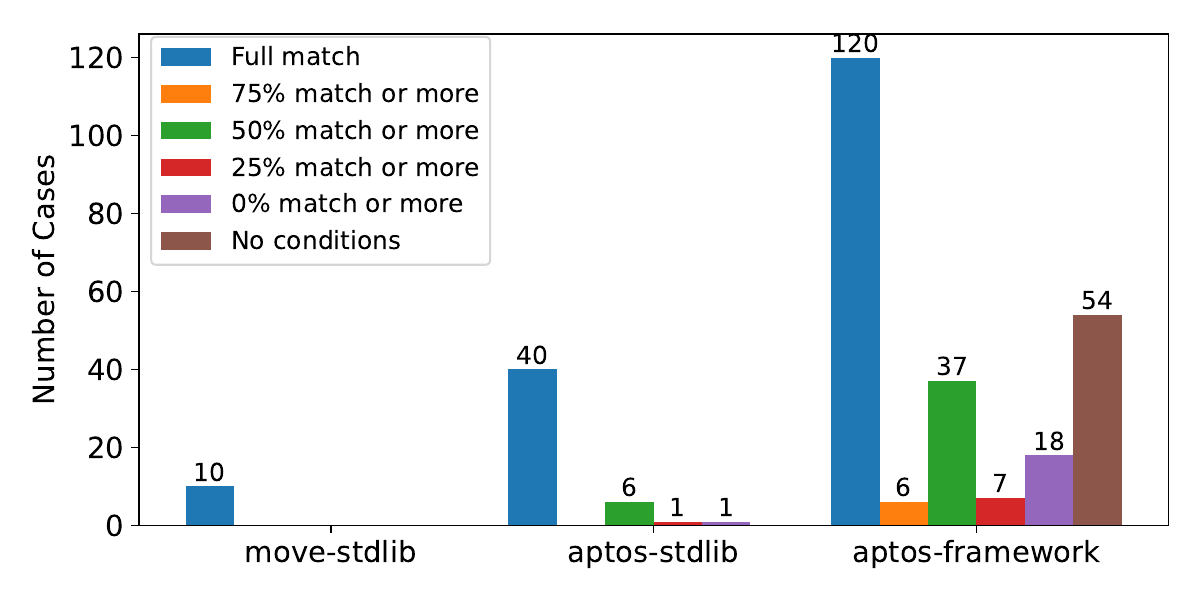}
  \caption{Specification comprehensiveness distribution
    of verifiable specifications by \sys.}
  \label{fig:spec-stats-o3-mini}
\end{figure}

\PP{Specification Comprehensiveness}
\autorefp{fig:spec-stats-o3-mini} shows the distribution of
specification comprehensiveness percentage
for all generated specifications by \sys that are verifiable
across the Aptos libraries.
A "full match" indicates that all manually written clauses in the codebase
(e.g., \cc{ensures}, \cc{aborts_if}, and \cc{modifies})
are implied by the generated specification.
\sys shows strong performance in the less complex libraries
(move-stdlib and aptos-stdlib) with 86.2\% (50/58) fully matched,
but achieves only 63.8\%
(120/188, excluding functions without ground-truth conditions)
in the more complex aptos-framework.
We observe that the missing clauses from \sys
are mostly \cc{aborts_if} clauses,
which LLMs tend to miss in functions with deeply-nested calls.

\begin{table}[ht]
  \centering
  \setlength{\tabcolsep}{3pt} %
  \begin{tabular}{c@{}ccc}
    \toprule
    & \textbf{move-stdlib} & \textbf{aptos-stdlib} & \textbf{aptos-framework} \\
    \midrule
    \textbf{Verified Functions}       & 10  & 48    & 242 \\
    \textbf{Expert-written Clauses}     & 29  & 149   & 442 \\
    \textbf{Generated Clauses} & 29  & 143   & 690 \\
    \textbf{Matched Clauses}     & 29  & 133   & 347 \\
    \textbf{Loop Invariants}         & 8  & 1   & 5 \\
    \midrule
    \midrule
    \textbf{Unique \cc{ensures}}     & 0  & 4   & 126 \\
    \textbf{Unique \cc{aborts\_if}}  & 1  & 3   & 82  \\
    \textbf{Unique \cc{modifies}}    & 0  & 3   & 72  \\
    \midrule
    \midrule
    \textbf{$\frac{\text{Match}}{\text{Total}}$} & 100\% & 89.3\% & 78.5\% \\
    \textbf{$\frac{\text{Generated}}{\text{Total}}$} & 100\% &
    96\% & 156.1\% \\
    \textbf{$\frac{\text{Uniques}}{\text{Generated}}$} & 3.4\% &
    7\% & 40.6\% \\
    \bottomrule
  \end{tabular}
  \caption{Comparison of the number of verified functions, expert-written,
    generated, matched clauses, loop invariants, and unique clauses for
    different codebases. Please refer to \autoref{tab:benchmark}: loop
      invariants are from 17 selected loop-containing functions.}
\label{tab:spec-stats-o3-mini-conditions}
\end{table}

\begin{table}[ht]
  \centering
  \begin{tabular}{lccccc}
    \toprule
    Variant & A & P & S & I & V \\
    \midrule
    \sys           & \cmark & \cmark & \cmark & \xmark & 84.0\%  \\
    \sysm          & \cmark & \xmark & \cmark & \xmark & 70.9\% \\
    \sysinline     & \cmark & \cmark & \cmark & \cmark & 83.8\%  \\
    \midrule
    \midrule
    \sysaio        & \xmark & \cmark & \cmark & \xmark & 53.5\%  \\
    \sysaiom       & \xmark & \xmark & \cmark & \xmark & 41.1\%  \\
    \sysaioinline  & \xmark & \cmark & \cmark & \cmark & 54.1\%  \\
    \sysaionaive   & \xmark & \xmark & \xmark & \xmark & 38.3\%  \\
    \bottomrule
  \end{tabular}
  \caption{
    Feature breakdown and verifiability for all variants used in the
    evaluation and ablation study.
    Abbreviations: A = Agentic design, P = Prover feedback, S = System Prompt \&
    Static Analysis, I = Function inlining, V = Verifiability rate with
    o3-mini.
  }
  \label{tab:variants}
\end{table}

\autoref{tab:spec-stats-o3-mini-conditions} summarizes the \emph{aggregated} statistics:
\sys generates
100\% clauses for move-stdlib,
89.3\% for aptos-stdlib, and
78.5\% for aptos-framework,
aggregating to 82.1\% overall.
Although some generated specifications are not fully matched, they remain
useful—often yielding even \emph{aggregately} more verification conditions than
the expert-written versions (\emph{aggregately} 139\%: 82\% matching ones plus
an additional 57\% that differ).
Among them, \sys also generates \emph{unique clauses} that manually-written
specifications miss:
3.4\% in move-stdlib,
7\% in aptos-stdlib,
40.6\% in aptos-framework, and overall
33.2\% (291) of all 876 generated clauses.
The excellent performance corresponds to less presence of specification
  coverage feedback discussed later in \autoref{ss:eval-cov}.

Overall, these statistics answer \textbf{RQ1}: LLMs can reason about Move
programs well even though Move is an emerging language, which is validated by
the strong performance of \sys in generating not only more verifiable
specifications but unique ones compared to expert-written specifications.

\PP{Reasons for Not Generating Specifications}
\sys fails to generate specifications for 57 functions
(or rather it generates an abstract specification as a last resort).
The main reasons include:
\begin{itemizesqn}
\item \emph{Calling impure functions.}
  Impure functions are Move functions that have side effects,
  such as modifying global storage and
  those functions should not be called
  as an expression in the specification.
  Instead, the idiomatic way is to write a pure version of the function
  as a helper function and call it in the specification.
  While we explicitly instruct the LLM to do this,
  it still sometimes calls them,
  causing failures.

\item \emph{Calling non-existent functions.}
  We observe that LLMs might still hallucinate and
  generate specifications that
  invoke non-existent functions.
  This could be due to the fact that LLMs are trained with
  limited data on Move and improvised with
  knowledge from other programming languages.

\item \emph{Other compiler errors.}
  LLMs sometimes generate code that is not valid Move syntax,
  including type errors or
  calling a function with the wrong number of arguments.
  This might be another sign of hallucination.

\item \emph{Complexity of the target function.}
  Some functions are too complex for the LLM to understand,
  especially those with deeply-nested calls or branches
  that appear in the aptos-framework
  (as shown in \autoref{fig:spec-stats-o3-mini} as well).
\end{itemizesqn}

\subsection{Ablation Study for \sys and \sysaio}
\label{ss:ablation-study}

In this section, we evaluate the agentic design and its underlying components.
This ablation study highlights the contributions of each component in our
design.

\PN{\sysaio: All-in-one Version of \sys.}
In addition to the agentic version of \sys, we implement an all-in-one variant,
\sysaio, which was our initial approach to explore the capability of LLMs in
generating Move specifications.
The key difference is that \sysaio generates all classes of specifications in a
single generation loop, whereas \sys generates them separately and then merges
the results using the specification ensembler.
Notably, \sysaio lacks the prover error summary agent, the guidance for
addressing prover errors, and the post-processing steps used to resolve common
prover issues.

To illustrate the benefits of our agentic design, we compare \sys with \sysaio.
Furthermore, to assess the individual contributions of the underlying components
(i.e., curated system prompts, static analysis, and prover feedback), we
evaluate three variants of \sysaio:
\begin{itemizesqn}
  \item \textbf{\sysaionaive:} A baseline version without any features.
  \item \textbf{\sysaiom:} \sysaionaive enhanced with a system prompt and static
    analysis, but without prover feedback.
  \item \textbf{\sysaio:} The full \sysaio incorporating the Move Prover.
\end{itemizesqn}
Based on our evaluation, prover feedback significantly contributes to the
overall accuracy of \sysaio.
Consequently, we introduce a variant of \sys, called \sysm, which employs the
agentic design without incorporating Move Prover feedback.
In later paragraphs, we evaluate the effect of function inlining with
  \sysinline and \sysaioinline.
\autoref{tab:variants} shows feature breakdown and verifiability for all
  variants that will be used in the evaluation and ablation study.

\autoref{tab:merged-results} presents the breakdown and aggregated results
  from three runs, comprising 357 functions with three trials per function.
In our evaluation framework, a function is considered successful if any one of
the three trials produces a verifiable specification.
If multiple trials succeed, the trial with the earliest successful round is
recorded.
\mainonly{Evaluation results for weaker models are in the extended
  version~\cite{msg-extended}. We selectively mentioned them in the following
  paragraphs.}

\PP{Effectiveness of the Agentic Design}
Comparing the results between \sys and \sysaio, the
benefits of the agentic design become clear.
With the advanced o3-mini model, \sys produced non-abstract specifications for
84\% of the cases (300 out of 357), while \sysaio achieved only 53.5\%.
In addition, \sys yielded abstract specifications for 57 functions, whereas
\sysaio generated 112 abstract specifications but outright failed on 54
functions.
These results demonstrate that the agentic design significantly enhances
verifiability, thereby addressing \textbf{RQ2}: leveraging MSL features (four
compositional classes of specifications) truly makes a difference.

\PP{Round and Move Prover Feedback}
To assess the impact of Move Prover feedback for \textbf{RQ3}, we compare
\sysaiom with \sysaio, and \sys with \sysm, respectively.
With additional rounds in the generation loop, the LLMs have more opportunities
to refine or fix the generated specifications with the aid of the Move Prover.
In contrast, without prover feedback, the later rounds offer limited corrective
value, as the LLM lacks guidance regarding specific errors.

The results show that Move Prover feedback is indeed beneficial for the
generation process.
For example, in the all-in-one design \sysaio, with o3-mini, 30.9\% (59 out of
191) of verifiable specifications were generated in later rounds using prover
feedback, compared to only 4.8\% in \sysaiom (high one-shot rate 95.2\%,
  but little self-correction).
Similarly, when comparing the agentic design \sys with \sysm, the overall
accuracy for \sysm drops by 13.1\% from 84\% to 70.9\% after removing prover
feedback.
Moreover, \sysm verifies only 13.1\% fewer functions while generating just
67.6\% of the clauses compared to \sys (a 32.4\% reduction), indicating that
Move Prover feedback improves both the accuracy and quality of generated
specifications.

This substantial difference demonstrates that the improvement primarily stems
from incorporating prover feedback rather than solely increasing the number of
rounds, thereby answering \textbf{RQ3}: Move Prover feedback is indeed useful.

\PP{System Prompts and Static Analysis} The naive version of all-in-one design,
\sysaionaive, achieves only 38.3\% verifiability with o3-mini and 18.4\% with
\mainarxiv{GPT-4o-mini (detailed results in extended version~\cite{msg-extended}),}{GPT-4o-mini (detailed results in \autoref{tab:ablation}),}
showing that without enriched prompts, the LLM generates incorrect Move syntax
and ignores callee function contents.
By using a curated system prompt with examples and applying static analysis to
capture function dependencies, \sysaiom significantly improves performance:
41.4\% verifiability with o3-mini and 29.1\% with GPT-4o-mini—a 58.2\%
improvement over \sysaionaive for GPT-4o-mini.
This demonstrates the value of these enhancements, particularly for less capable
LLMs, and shows that static analysis effectively builds the correct context for
specification generation, partially answering \textbf{RQ5}.

\PP{Best-effort Function Inlining}
We assess whether inlining functions affects \sys's performance, which
ultimately depends on the LLM's understanding of function calls.
Inlining is a double-edged sword: it reduces the number of functions processed,
but may increase the target function's complexity—especially when the inlined
function contains branches (e.g., if-else statements).
\autoref{tab:merged-results} includes results for the normal (with
  conversation context \textbf{V1}) and inlined versions (\textbf{V1} and
  \textbf{V2}) of \sys and \sysaio: \sysinline and \sysaioinline.
Overall, the aggregated performance difference is minimal: \sys generates one
more specification than \sysinline, and \sysaio generates two fewer than
\sysaioinline.

\begin{table}[ht]
  \centering
  \begin{tabular}{ccccc}
    \toprule
    & Normal-only & Common & Inline-only &\\
    \midrule
    \sys     &   8      &  292   &   7    & \sysinline  \\
    \sysaio &   18      & 173       & 20  & \sysaioinline      \\
    \bottomrule
  \end{tabular}
  \caption{Table representation of Venn diagrams comparing the generated
      specifications between the normal and inlined variants for \sys and
      \sysaio.
      ``Common'' means the specifications that could be found by both normal and
      inlined versions. }
  \label{tab:inline-venn-table}
\end{table}

\autorefp{tab:inline-venn-table} further uses Venn
  diagram~\cite{venn-diagram} (in table form) to illustrate the specification
sets produced by the normal and inlined versions for both \sys and \sysaio.
The evaluation results confirm our hypothesis: inlining has both
positive and negative effects on the LLM's specification-generation capability.
Specifically, \sysinline generates 7 specifications that are missed by \sys,
while it misses 8 specifications that \sys does produce.
Similarly, \sysaioinline generates 20 specifications not produced by \sysaio,
but fails to generate 18 ones that \sysaio does.

Analysis of these discrepancies reveals that the inlined version could be
overwhelmed by the complexity of inlined functions (e.g. deeply-nested calls or
branches).
In such cases, attempting to cover all aspects of the function in a single large
body could fail, even with the corrective help of Move Prover.

On the positive side, we observe that the inlined version sometimes offers
better reasoning than the normal version, particularly when dealing with
\emph{reasonably-nested} calls.
In scenarios where multiple functions with small bodies are combined into a
single larger function, the unified context can be easier for the LLM to
understand, which highlights our design choice for best-effort function
inlining.

In conclusion, our results confirm that inlining is beneficial for certain
programs, which partially addresses \textbf{RQ4}.
A promising approach to optimize performance is to combine the advantages of
both the normal and inlined versions.
Running them in parallel and subsequently selecting or merging the best results
using a specification ensembler can lead to overall improved generation
verifiability.

\PP{Summary}
Our ablation study shows that the agentic design of \sys significantly improves
accuracy by leveraging the compositional nature of various Move specifications
(\textbf{RQ2}).
With Move Prover feedback, both \sys and \sysaio gain the ability to self-fix
erroneous specifications, drastically improving accuracy and quality
(\textbf{RQ3}).
Finally, scaffolding utilities like static analysis and best-effort function
inlining effectively enhance the accuracy of generated specifications by
providing well-scoped conversation contexts (\textbf{RQ4}).

\subsection{Comprehensiveness for Generated Specifications}
\label{ss:eval-cov}
To evaluate the comprehensiveness of the generated specifications, we utilize
the specification coverage component in \sys.
During the evaluation, the specification coverage component was enabled in the
\cc{ensures} ClausesGen agent, which attempts to identify missing \cc{ensures}
clauses, similar to those illustrated in the simplified example in
\autoref{fig:example-random-deletion}.
By examining the execution traces of \sys and \sysaio, we filter cases where
specification coverage feedback is triggered (i.e., non-empty).
Using \texttt{o3-mini}, we conducted total \num{1071} trials (\ie, 357
  functions with 3 trials each in evaluation). Among these, only 66 trials for
  \sys and 77 trials for \sysaio triggered specification-coverage feedback.
This is primarily due to the fact that
the Move Prover helps \sys generate sufficiently complete specifications to
pass verification, which is the prerequisite for specification coverage
computation.
As there are a non-trivial number of unique clauses (specifically, unique
\cc{ensures} clauses) that \sys generates as shown in
\autoref{tab:spec-stats-o3-mini-conditions}, it's not surprising that generated
specifications are comprehensive enough, which leads to few observations of
specification coverage feedback.

For most of the observed cases, the feedback was triggered by the deletion of an
isolated assertion statement.
Since the LLM can capture the intention behind such assertions very well, as
it's a common feature for most programming languages, the LLM typically already
generates specifications that inherently cover these cases.

Thus, while specification coverage did not lead to noticeable improvements in
our current evaluation, its presence and traces confirm that \sys already
produces high-quality, comprehensive specifications.
Moreover, the use of specification coverage to evaluate comprehensiveness
of specifications is \emph{fully-automatic} without the need to manually compare
generated specifications with expert-written ones as we conduct in
\autoref{ss:eval-result}.
The results answer \textbf{RQ5}: non-trivial specification coverage is not
observed for \emph{verifiable} specifications generated by \sys and \sysaio: the
generated verifiable specifications are comprehensive enough when they
pass the Move Prover.

\section{Discussion}
\label{s:discussion}
\subsection{Adaptation to Other Verification Languages}
\label{ss:adaptation}

\sys's \emph{language-agnostic} approach combines compositional generation with
an automated specification-coverage metric to ensure comprehensiveness.

\PP{Compositional Generation} Compositional generation scales by isolating
orthogonal property groups such as post-conditions, aborts (which serve similar
purposes to pre-conditions\footnote{actual pre-conditions specified by
  \cc{require} appear in just $\approx2\%$ of Aptos functions—typically to
  assume external assumptions (\eg., "blockchain is running").} in
other languages), global-state updates in Move for blockchains.
The approach can be adapted to other languages (\eg, Dafny~\cite{dafny},
Why3~\cite{why3} for general programs, or solc-verify~\cite{solc-verify},
Certora~\cite{certora} for Solidity~\cite{solidity} used to verify
Ethereum~\cite{ethereum} smart contracts) by grouping properties around
domain-specific concerns.

\PP{Certora}
For example, a straightforward adoption is verifying contracts in Ethereum with
Certora Prover.
Due to the different natures of Move/Ethereum and Move Prover/Certora
Prover, the specific syntax/implementation might be different;
however, these blockchain-specific concerns are shared between both platforms
(\ie, abortion/revert, global-state changes).
To be more concrete, in Certora Verification Language (CVL),
we could combine
\cc{require} (pre-condition), \cc{fun@withrevert} annotation
(asks the prover to set \cc{lastReverted} based on
whether the function \cc{fun} reverted), and
\cc{assert lastReverted} to test
whether a function will abort under certain conditions, which corresponds to
\cc{aborts\_if} in MSL.
Testing global-state updates requires more scaffolding due to the lack of
\cc{modifies} clauses;
this can be addressed by using ghost variables~\cite{certora-ghosts} to record
additional variable updates during verification, after which the modification of
final states can be checked with \cc{assert} by comparing with original states.
As long as the verification languages have enough constructs (\cc{withrevert},
and ghost variables in Certora, for example), the adaptation will be easy and
straightforward for other blockchain smart contract verifiers.

For other domains (\eg, operating systems, network protocols), it requires the
experts to identify distinct groups of domain-specific concerns that could
be verified independently.

\PP{Specification Coverage}
Specification coverage only requires language-specific mutators for
random deletion used in \autoref{ss:spec-cov}.

\subsection{Threats to Validity: Data Exposure to LLM}
\label{ss:threat}

LLMs show strong generalization across domains and a solid grasp of Move,
despite its limited training data.
However, our evaluation shows that baseline \sysaionaive with advanced o3-mini
model proves only 38.3\% of specifications without advanced context engineering
and an agentic approach.
Thus, while familiarity with mainstream languages aids general capabilities, it
does not necessarily ensure strong performance in niche languages like Move,
especially for specification generation without proper system design.

\section{Related Works: LLM-based Specification}
\label{s:relwk}

As briefly discussed in~\autoref{s:intro},
\sys is built upon the same insights highlighted in the recent
LLM for specification generation
works~\cite{autospec, specgen, xie2025effective,
osvbench, dafny-loop-gen,
liu2024enhancing, kamath2023finding, chakraborty2023ranking, autoverus, SAFE,
lahirie2024evaluating, kamath2024leveraging}:
heuristic or rule-based specifications synthesis templates~\cite{houdini,
daikon} can hardly match the code comprehension and reasoning capabilities of
modern LLMs.
And yet, \sys provides more insights on generality of this research direction,
while also independently rediscovering and validating parts of previous
  works in low-resource languages such as Move:
\WC{1} LLM for specification generation has the potential to
generalize to even non-mainstream programing languages (e.g., Move);
\WC{2} an LLM-based specification generator should leverage
features in the underlying specification language at the design stage
(e.g., a separation of concerns of different clauses in MSL enables \sys to
generate specifications in a modular and agentic way);
\WC{3} leveraging feedback, especially
counterexamples~\cite{kamath2024leveraging},
from the verification toolchain (not only the compiler) can significantly
improve the quality of generated specifications;
\WC{4} simple metrics such as specification coverage can be used to
gauge the comprehensiveness of generated specifications;
\WC{5} scaffolding utilities
(e.g., static analysis tools or function inliners)
can be helpful to scoping the context~\cite{shrivastava2023repository} for LLMs
to generate specifications,
but the engineering effort to build such utilities
should not be underestimated.
We hope the experience and insights from \sys can
inspire future works in specification generation and
LLM4FM research in general.
We highlight key differences with two particularly relevant works:

\PN{\autoverus.} \phantomsection\label{rev-autoverus}
\autoverus~\cite{autoverus} uses specialized repair agents, each addressing a
specific prover error, based on the idea that specification repairing can be
decomposed into sub-problems aligned with common failure modes.
They implemented
10 such agents with tailored prompts.
Similarly, \sys leverages the compositional
structure of the Move Specification Language, assigning different prompts to
different specification parts in an integrated way.
The two approaches are
orthogonal and could be fruitfully combined.
Notably, besides specialized
prompts for different classes of clauses, \sys already uses specialized prompts
for some prover errors, but does not (yet) isolate them into separate agents as
\autoverus does.

\PP{Testcase-based Evaluation of Specifications}
Testcase-based evaluation used in \cite{lahirie2024evaluating} helps gauge
specification comprehensiveness through differential testing, where both buggy
and correct implementations are checked against the specification to see whether
it captures the intended behavior.
However, this approach relies on the
availability of such paired implementations, which are often scarce in emerging
languages like Move.
Our specification coverage metric is implementation-agnostic, making it
particularly useful where suitable code versions are lacking.
It assesses specification quality via code mutation (random deletion) and
verifiability.
Both approaches are complementary, and ideally should be used together for a
thorough evaluation.

\section{Conclusion}
\label{s:conclusion}
In this paper, we present \sys, a fully automated, end-to-end Move specification
generation agent for Move smart contracts.
Building on an agentic design that leverages the compositional characteristics
of the Move Specification Language, \sys integrates several scaffolding
utilities, including static analysis, best-effort function inlining for a
well-scoped LLM conversation context, Move Prover feedback as a
verifier-in-the-loop oracle, and a simple metric---specification coverage---to
automatically evaluate the comprehensiveness of the generated specifications
during the generation process.
Evaluation results demonstrate that \sys not only produces verifiable
expert-level specifications but also generates unique specifications, proving
its capability to ease the burden of formal verification for Move smart
contracts.

\mainarxiv{
  \newline
  \PP{Acknowledgment}%
}
{
\section*{Acknowledgment} %
\label{s:ack}%
}
\arxivonly{We thank the anonymous reviewers for their insightful feedback.}
This research was supported by Sui Foundation, COYA INNOVATION and ONR
under grant N00014-23-1-2095.

\bibliographystyle{IEEEtran}
\bibliography{p,sslab,conf}

\begin{thebibliography}{10}
\providecommand{\url}[1]{#1}
\csname url@samestyle\endcsname
\providecommand{\newblock}{\relax}
\providecommand{\bibinfo}[2]{#2}
\providecommand{\BIBentrySTDinterwordspacing}{\spaceskip=0pt\relax}
\providecommand{\BIBentryALTinterwordstretchfactor}{4}
\providecommand{\BIBentryALTinterwordspacing}{\spaceskip=\fontdimen2\font plus
\BIBentryALTinterwordstretchfactor\fontdimen3\font minus
  \fontdimen4\font\relax}
\providecommand{\BIBforeignlanguage}[2]{{%
\expandafter\ifx\csname l@#1\endcsname\relax
\typeout{** WARNING: IEEEtran.bst: No hyphenation pattern has been}%
\typeout{** loaded for the language `#1'. Using the pattern for}%
\typeout{** the default language instead.}%
\else
\language=\csname l@#1\endcsname
\fi
#2}}
\providecommand{\BIBdecl}{\relax}
\BIBdecl

\bibitem{defipulse2022}
{Concourse Open Community}, ``{DeFi Pulse},'' \url{https://defipulse.com/},
  2022, accessed: Mar 1, 2025.

\bibitem{cryptosec2022}
{CryptoSec Group}, ``Documented timeline of defi exploits,''
  \url{https://cryptosec.info/defi-hacks/}, 2022, accessed: Mar 1, 2025.

\bibitem{clarke1999model}
E.~M. Clarke, O.~Grumberg, and D.~A. Peled, \emph{Model Checking}.\hskip 1em
  plus 0.5em minus 0.4em\relax MIT Press, 1999.

\bibitem{bertot2004interactive}
Y.~Bertot and P.~Cast{\'e}ran, \emph{Interactive Theorem Proving and Program
  Development: Coq'Art: The Calculus of Inductive Constructions}, ser. Texts in
  Theoretical Computer Science. An EATCS Series.\hskip 1em plus 0.5em minus
  0.4em\relax Springer, 2004.

\bibitem{felderer2018formal}
M.~Felderer, D.~Gurov, M.~Huisman, B.~Lisper, and R.~Schlick, ``Formal methods
  in industrial practice - bridging the gap (track summary),'' in
  \emph{Leveraging Applications of Formal Methods, Verification and Validation.
  Industrial Practice}, Rhodes, Greece, Oct. 2018.

\bibitem{ferrari2022formal}
A.~Ferrari and M.~H.~T. Beek, ``Formal methods in railways: A systematic
  mapping study,'' \emph{ACM Computing Surveys}, vol.~55, no.~4, Nov. 2022.

\bibitem{gleirscher2020formal}
M.~Gleirscher and D.~Marmsoler, ``Formal methods in dependable systems
  engineering: A survey of professionals from europe and north america,''
  \emph{Empirical Software Engineering}, vol.~25, no.~6, 2020.

\bibitem{kulik2022survey}
T.~Kulik, B.~Dongol, P.~G. Larsen, H.~D. Macedo, S.~Schneider, P.~W.~V.
  Tran-J{\o}rgensen, and J.~Woodcock, ``A survey of practical formal methods
  for security,'' \emph{Formal Aspects of Computing}, vol.~34, no.~1, Jul.
  2022.

\bibitem{certora-report1}
{Certota Team}, ``Certora securiy report - mayan fastmctp,''
  \url{https://www.certora.com/reports/mayan-smart-contract-security-report},
  2025, accessed: May, 2025.

\bibitem{certora-report2}
------, ``Certora securiy report -
  https://www.certora.com/reports/texture-finance-security-report,''
  \url{https://www.certora.com/reports/mayan-smart-contract-security-report},
  2025, accessed: May, 2025.

\bibitem{autospec}
C.~Wen, J.~Cao, J.~Su, Z.~Xu, S.~Qin, M.~He, H.~Li, S.-C. Cheung, and C.~Tian,
  ``Enchanting program specification synthesis by large language models using
  static analysis and program verification,'' in \emph{International Conference
  on Computer Aided Verification}.\hskip 1em plus 0.5em minus 0.4em\relax
  Springer, 2024, pp. 302--328.

\bibitem{specgen}
\BIBentryALTinterwordspacing
L.~Ma, S.~Liu, Y.~Li, X.~Xie, and L.~Bu, ``{ SpecGen: Automated Generation of
  Formal Program Specifications via Large Language Models },'' in \emph{2025
  IEEE/ACM 47th International Conference on Software Engineering (ICSE)}.\hskip
  1em plus 0.5em minus 0.4em\relax Los Alamitos, CA, USA: IEEE Computer
  Society, May 2025, pp. 666--666. [Online]. Available:
  \url{https://doi.ieeecomputersociety.org/10.1109/ICSE55347.2025.00129}
\BIBentrySTDinterwordspacing

\bibitem{xie2025effective}
D.~Xie, B.~Yoo, N.~Jiang, M.~Kim, L.~Tan, X.~Zhang, and J.~S. Lee, ``How
  effective are large language models in generating software specifications,''
  \emph{arXiv preprint arXiv:2306.03324}, 2025.

\bibitem{osvbench}
\BIBentryALTinterwordspacing
S.~Li, J.~Jiang, T.~Zhao, and J.~Shen, ``Osvbench: Benchmarking llms on
  specification generation tasks for operating system verification,'' 2025.
  [Online]. Available: \url{https://arxiv.org/abs/2504.20964}
\BIBentrySTDinterwordspacing

\bibitem{dafny-loop-gen}
J.~Pascoal~Faria, E.~Trigo, and R.~Abreu, ``Automatic generation of loop
  invariants in dafny with large language models,'' in \emph{International
  Conference on Fundamentals of Software Engineering}.\hskip 1em plus 0.5em
  minus 0.4em\relax Springer, 2025, pp. 138--154.

\bibitem{liu2024enhancing}
R.~Liu, G.~Li, M.~Chen, L.-I. Wu, and J.~Ke, ``Enhancing automated loop
  invariant generation for complex programs with large language models,''
  \emph{arXiv preprint arXiv:2412.10483}, 2024.

\bibitem{kamath2023finding}
A.~Kamath, A.~Senthilnathan, S.~Chakraborty, P.~Deligiannis, S.~K. Lahiri,
  A.~Lal, A.~Rastogi, S.~Roy, and R.~Sharma, ``Finding inductive loop
  invariants using large language models. corr abs/2311.07948 (2023),'' 2023.

\bibitem{chakraborty2023ranking}
S.~Chakraborty, S.~K. Lahiri, S.~Fakhoury, M.~Musuvathi, A.~Lal, A.~Rastogi,
  A.~Senthilnathan, R.~Sharma, and N.~Swamy, ``Ranking llm-generated loop
  invariants for program verification,'' \emph{arXiv preprint
  arXiv:2310.09342}, 2023.

\bibitem{liu2024propertygpt}
Y.~Liu, Y.~Xue, D.~Wu, Y.~Sun, Y.~Li, M.~Shi, and Y.~Liu, ``Propertygpt:
  Llm-driven formal verification of smart contracts through retrieval-augmented
  property generation,'' \emph{arXiv preprint arXiv:2405.02580}, 2024.

\bibitem{sun2024gptscan}
Y.~Sun, D.~Wu, Y.~Xue, H.~Liu, H.~Wang, Z.~Xu, X.~Xie, and Y.~Liu, ``Gptscan:
  Detecting logic vulnerabilities in smart contracts by combining gpt with
  program analysis,'' in \emph{Proceedings of the IEEE/ACM 46th International
  Conference on Software Engineering}, 2024, pp. 1--13.

\bibitem{houdini}
C.~Flanagan and K.~R.~M. Leino, ``Houdini, an annotation assistant for
  {ESC}/{Java},'' in \emph{Formal Methods Europe (FME 2001): Formal Methods for
  Increasing Software Productivity}, ser. Lecture Notes in Computer Science,
  vol. 2021.\hskip 1em plus 0.5em minus 0.4em\relax Springer, 2001, pp.
  500--517.

\bibitem{daikon}
M.~D. Ernst, J.~H. Perkins, P.~J. Guo, S.~McCamant, C.~Pacheco, M.~S. Tschantz,
  and C.~Xiao, ``The {Daikon} system for dynamic detection of likely
  invariants,'' in \emph{Science of Computer Programming}, vol.~69, no.
  1--3.\hskip 1em plus 0.5em minus 0.4em\relax Elsevier, 2007, pp. 35--45.

\bibitem{autoverus}
C.~Yang, X.~Li, M.~R.~H. Misu, J.~Yao, W.~Cui, Y.~Gong, C.~Hawblitzel,
  S.~Lahiri, J.~R. Lorch, S.~Lu \emph{et~al.}, ``Autoverus: Automated proof
  generation for rust code,'' \emph{arXiv preprint arXiv:2409.13082}, 2024.

\bibitem{SAFE}
\BIBentryALTinterwordspacing
T.~Chen, S.~Lu, S.~Lu, Y.~Gong, C.~Yang, X.~Li, M.~R.~H. Misu, H.~Yu, N.~Duan,
  P.~Cheng, F.~Yang, S.~K. Lahiri, T.~Xie, and L.~Zhou, ``Automated proof
  generation for rust code via self-evolution,'' 2025. [Online]. Available:
  \url{https://arxiv.org/abs/2410.15756}
\BIBentrySTDinterwordspacing

\bibitem{lahirie2024evaluating}
S.~K. Lahirie, ``Evaluating llm-driven user-intent formalization for
  verification-aware languages,'' in \emph{2024 Formal Methods in
  Computer-Aided Design (FMCAD)}.\hskip 1em plus 0.5em minus 0.4em\relax IEEE,
  2024, pp. 142--147.

\bibitem{kamath2024leveraging}
A.~Kamath, N.~Mohammed, A.~Senthilnathan, S.~Chakraborty, P.~Deligiannis, S.~K.
  Lahiri, A.~Lal, A.~Rastogi, S.~Roy, and R.~Sharma, ``Leveraging llms for
  program verification,'' in \emph{2024 Formal Methods in Computer-Aided Design
  (FMCAD)}.\hskip 1em plus 0.5em minus 0.4em\relax IEEE, 2024, pp. 107--118.

\bibitem{shrivastava2023repository}
D.~Shrivastava, H.~Larochelle, and D.~Tarlow, ``Repository-level prompt
  generation for large language models of code,'' in \emph{International
  Conference on Machine Learning}.\hskip 1em plus 0.5em minus 0.4em\relax PMLR,
  2023, pp. 31\,693--31\,715.

\bibitem{move}
S.~Blackshear, E.~Cheng, D.~L. Dill, V.~Gao, B.~Maurer, T.~Nowacki, A.~Pott,
  S.~Qadeer, Rain, D.~Russi, S.~Sezer, T.~Zakian, and R.~Zhou, ``Move: A
  language with programmable resources,''
  \url{https://diem-developers-components.netlify.app/papers/diem-move-a-language-with-programmable-resources/2020-05-26.pdf},
  2020, accessed: 2025-03-01.

\bibitem{diem}
D.~Association, ``Diem blockchain: A scalable and secure blockchain for the
  digital economy,'' \url{https://www.diem.com/}, 2019, originally launched as
  Libra. Accessed: April 1, 2025.

\bibitem{aptos}
{Aptos Foundation}, ``Aptos,'' \url{https://aptoslabs.com}, 2022, accessed:
  Feb, 2025.

\bibitem{sui}
{Sui Foundation}, ``Sui,'' \url{https://sui.io}, 2022, accessed: Feb, 2025.

\bibitem{movement}
{Movement Network Foundation}, ``Movement,''
  \url{https://www.movementnetwork.xyz}, 2025, accessed: May, 2025.

\bibitem{move-prover}
D.~Dill, W.~Grieskamp, J.~Park, S.~Qadeer, M.~Xu, and E.~Zhong, ``Fast and
  reliable formal verification of smart contracts with the move prover,'' in
  \emph{Proceedings of the 28th International Conference on Tools and
  Algorithms for the Construction and Analysis of Systems (TACAS)}, Munich,
  Germany, Apr. 2022.

\bibitem{smt}
C.~Tinelli and C.~Barrett, ``Satisfiability modulo theories,'' in
  \emph{Handbook of Model Checking}, E.~M. Clarke, T.~A. Henzinger, H.~Veith,
  and R.~Bloem, Eds.\hskip 1em plus 0.5em minus 0.4em\relax Springer, 2018, pp.
  305--343.

\bibitem{z3}
L.~de~Moura and N.~Bj{\o}rner, ``Z3: An efficient smt solver,'' in \emph{Tools
  and Algorithms for the Construction and Analysis of Systems}, ser. Lecture
  Notes in Computer Science, C.~Ramakrishnan and J.~Rehof, Eds., vol.
  4963.\hskip 1em plus 0.5em minus 0.4em\relax Springer, 2008, pp. 337--340.

\bibitem{cvc5}
H.~Barbosa, C.~Barrett, M.~Brain, G.~Kremer, H.~Lachnitt, M.~Mann, A.~Mohamed,
  M.~Mohamed, A.~Niemetz, A.~N{\"o}tzli \emph{et~al.}, ``cvc5: A versatile and
  industrial-strength smt solver,'' in \emph{International Conference on Tools
  and Algorithms for the Construction and Analysis of Systems}.\hskip 1em plus
  0.5em minus 0.4em\relax Springer, 2022, pp. 415--442.

\bibitem{hoare1969axiomatic}
C.~A.~R. Hoare, ``An axiomatic basis for computer programming,''
  \emph{Communications of the ACM}, vol.~12, no.~10, pp. 576--580, 1969.

\bibitem{boogie}
M.~Barnett, B.-Y.~E. Chang, R.~DeLine, B.~Jacobs, and K.~R.~M. Leino, ``Boogie:
  A modular reusable verifier for object-oriented programs,'' Microsoft
  Research, Technical Report MSR-TR-2005-70, 2005,
  \url{https://www.microsoft.com/en-us/research/publication/boogie-a-modular-reusable-verifier-for-object-oriented-programs/}.

\bibitem{agent-reflexion}
\BIBentryALTinterwordspacing
N.~Shinn, F.~Cassano, A.~Gopinath, K.~R. Narasimhan, and S.~Yao, ``Reflexion:
  language agents with verbal reinforcement learning,'' in \emph{Thirty-seventh
  Conference on Neural Information Processing Systems}, 2023. [Online].
  Available: \url{https://openreview.net/forum?id=vAElhFcKW6}
\BIBentrySTDinterwordspacing

\bibitem{fast}
\BIBentryALTinterwordspacing
R.~Ji and M.~Xu, ``{ Finding Specification Blind Spots via Fuzz Testing },'' in
  \emph{2023 IEEE Symposium on Security and Privacy (SP)}.\hskip 1em plus 0.5em
  minus 0.4em\relax Los Alamitos, CA, USA: IEEE Computer Society, May 2023, pp.
  2708--2725. [Online]. Available:
  \url{https://doi.ieeecomputersociety.org/10.1109/SP46215.2023.10179438}
\BIBentrySTDinterwordspacing

\bibitem{aptos:abstract-spec}
{Aptos Foundation}, ``{Move Specification Language (Abstract Specification)},''
  \url{https://aptos.dev/build/smart-contracts/prover/spec-lang#abstract-specifications},
  2024, accessed: 2025-08-25.

\bibitem{uninterpreted-function}
{Wikipedia contributors}, ``Uninterpreted function,''
  \url{https://en.wikipedia.org/wiki/Uninterpreted_function}, 2023, [Online;
  accessed April 1, 2025].

\bibitem{subsumption}
G.~D. Plotkin, ``A note on inductive generalization,'' in \emph{Machine
  Intelligence 5}, B.~Meltzer and D.~Michie, Eds.\hskip 1em plus 0.5em minus
  0.4em\relax Edinburgh: Edinburgh University Press, 1970, pp. 153--163.

\bibitem{venn-diagram}
J.~Venn, ``I. on the diagrammatic and mechanical representation of propositions
  and reasonings,'' \emph{The London, Edinburgh, and Dublin Philosophical
  Magazine and Journal of Science}, vol.~10, no.~59, pp. 1--18, 1880.

\bibitem{dafny}
\BIBentryALTinterwordspacing
K.~R.~M. Leino, ``Dafny: An automatic program verifier for functional
  correctness,'' in \emph{Proceedings of the 16th International Conference on
  Logic for Programming, Artificial Intelligence, and Reasoning (LPAR-16)},
  ser. Lecture Notes in Computer Science, vol. 6355.\hskip 1em plus 0.5em minus
  0.4em\relax Springer, 2010, pp. 348--370. [Online]. Available:
  \url{https://link.springer.com/chapter/10.1007/978-3-642-17511-4_20}
\BIBentrySTDinterwordspacing

\bibitem{why3}
J.-C. Filli{\^a}tre and A.~Paskevich, ``Why3—where programs meet provers,''
  in \emph{European symposium on programming}.\hskip 1em plus 0.5em minus
  0.4em\relax Springer, 2013, pp. 125--128.

\bibitem{solc-verify}
{\'A}.~Hajdu and D.~Jovanovi{\'c}, ``solc-verify: A modular verifier for
  solidity smart contracts,'' in \emph{Working conference on verified software:
  theories, tools, and experiments}.\hskip 1em plus 0.5em minus 0.4em\relax
  Springer, 2019, pp. 161--179.

\bibitem{certora}
Certora, ``Certora: Formal verification for smart contracts,''
  \url{https://www.certora.com/}, accessed: 2025-08-19.

\bibitem{solidity}
{Solidity Contributors}, ``Solidity: A smart contract-oriented programming
  language,'' \url{https://soliditylang.org/}, 2014, version accessed: April 1,
  2025.

\bibitem{ethereum}
V.~Buterin \emph{et~al.}, ``Ethereum: A decentralized platform for
  applications,'' \url{https://ethereum.org/}, 2015, accessed: April 1, 2025.

\bibitem{certora-ghosts}
{Certora Ltd.}, ``Ghosts in {Certora} verification language,''
  \url{https://docs.certora.com/en/latest/docs/cvl/ghosts.html}, 2024,
  accessed: 2025-08-19.

\end{thebibliography}

\ifarxiv
\clearpage
\onecolumn
\appendix

\subsection{Abstract Specification Example}
\label{app:abstract-spec-example}

\begin{figure}[ht]
  \centering
  \input{code/abstract-spec-example.move}
  \caption{Abstract specification example for modular exponentiation function.}
  \label{fig:abstract-spec-example}
\end{figure}

\autoref{fig:abstract-spec-example} shows an abstract specification for a
complex mathematical function—modular exponentiation with binary
exponentiation—which is beyond the capabilities of the Move Prover due to
limitations in the underlying SMT solvers.
By specifying the behavior abstractly using ``[abstract]'', we can refer to the
desired mathematical result without providing a fully-verified implementation.
This approach also allows developers to use a provisional specification as a
placeholder until a concrete, prover-friendly version can be written, either
through approximation or with further tool improvements.
We provide a real example generated by \sys for Aptos in the case studies
(\autoref{ss:case-study}).

\subsection{Case Studies}
\label{ss:case-study}
To provide concrete insights into \sys's specification generation capabilities,
we present a detailed analysis of selected examples that
illustrate the system's operational mechanisms
and demonstrate its practical effectiveness.
Specifically, we examine two representative instances where
Move Prover feedback enables iterative refinement of
initially imperfect specifications,
along with a third case that exemplifies
how \sys can autonomously draft sophisticated specifications
for mathematically complex functions,
thereby providing valuable scaffolding for subsequent expert review and refinement.

\begin{figure*}[hbt]
  \centering
  \begin{subfigure}[t]{0.32\linewidth}
    \input{code/case-miss-abort.move}
    \caption{Move Function}
    \label{fig:case-miss-abort-fun}
  \end{subfigure}
  \hfill
  \begin{subfigure}[t]{0.32\linewidth}
    \input{code/case-miss-abort-missing.spec.move}
    \caption{Partial specification missing an abort clause}
    \label{fig:case-miss-abort-spec-missing}
  \end{subfigure}
  \hfill
  \begin{subfigure}[t]{0.32\linewidth}
    \input{code/case-miss-abort-correct.spec.move}
    \caption{Corrected full specification}
    \label{fig:case-miss-abort-spec-correct}
  \end{subfigure}
  \caption{Demonstration of iterative specification completeness enhancement
    through systematic incorporation of Move Prover diagnostics.
  }
  \label{fig:case-miss-abort}
\end{figure*}

\PN{Iterative Refinement of Missing \cc{aborts_if} Clauses.}
The empirical results presented in \autoref{tab:merged-results}
demonstrate the substantial utility of
Move Prover feedback in facilitating specification correctness
through systematic error detection and remediation.
\autoref{fig:case-miss-abort} presents a particularly clear example
where the synergistic interaction between LLM reasoning and
formal verification feedback enables the identification and
subsequent correction of an omitted \cc{abort\_if} clause.
The function under consideration exhibits abort behavior
when the requisite \cc{ObjectCore} resource is absent
from the blockchain's global storage at the address specified by \cc{ref.self}.
In the initial generation phase, \sys produces a specification
(\autoref{fig:case-miss-abort-spec-missing})
that, while syntactically well-formed,
omits the crucial \cc{aborts\_{if}} clause corresponding to this failure condition,
consequently rendering the specification unverifiable by the Move Prover.
The diagnostic output generated by the Move Prover
(with \cc{o} representing \cc{object}, abbreviated for space)
illustrates this deficiency:
\begin{mdframed}[linewidth=1pt]
  \begin{minipage}{\linewidth}
    \footnotesize
\begin{verbatim}
error:abort not covered by any `aborts_if` clauses
|
|let o = borrow_global_mut<ObjectCore>(ref.self);
|        -- abort happened here with execution failure
\end{verbatim}
  \end{minipage}
\end{mdframed}
This diagnostic precisely identifies the unhandled abort condition
that was inadvertently omitted from the initial specification.
Through the incorporation of this formal verification feedback
into subsequent generation iterations,
the LLM demonstrates remarkable capacity for
error recognition and specification refinement,
ultimately producing the corrected specification
(\autoref{fig:case-miss-abort-spec-correct})
that comprehensively addresses the identified deficiency.

\begin{figure}[ht]
  \centering
  \input{code/create-object.move}
  \caption{Example of autonomous postcondition refinement through formal verification feedback.}
  \label{fig:fix-ensure}
\end{figure}

\PN{Automated Correction of Erroneous \cc{ensures} Clauses.}
\autoref{fig:fix-ensure} exemplifies the sophisticated self-correction capabilities
of \sys when confronted with semantically incorrect postcondition specifications,
demonstrating the system's ability to leverage formal verification feedback
for iterative specification refinement.
The \cc{create\_object} function implements a complex protocol for
object instantiation within the blockchain's global storage namespace,
ensuring address uniqueness through careful resource management.
The core functionality is encapsulated within \cc{create\_object\_internal} (Line 7),
which performs a multi-stage creation process.
Initially, the function constructs a signer object---a cryptographic primitive
essential for authenticated modifications to account-specific global storage.
Subsequently, at Line 12, the function invokes \cc{guid::create}
to generate a globally unique identifier (\cc{transfer\_events\_guid}),
passing a \emph{mutable reference} to \cc{guid\_creation\_num}
initialized with the constant \cc{INIT\_GUID\_CREATION\_NUM}.
Crucially, \cc{guid::create} exhibits side-effect behavior:
it returns the original value of \cc{guid\_creation\_num}
while simultaneously incrementing the referenced value,
thereby ensuring subsequent GUID uniqueness.
The function completes by creating an \cc{ObjectCore} object
parameterized by the generated GUID,
returning a \cc{ConstructorRef} handle for further manipulation.
The figure presents two specifications generated by \sys
(with extraneous elements elided for clarity),
which illuminate the system's iterative refinement process.
The initial specification (Line 30) exhibits a subtle but critical semantic flaw:
it fails to properly model the side-effect induced by
the mutable reference semantics of \cc{guid\_creation\_num}.
This oversight manifests in the Move Prover's diagnostic output
(presenting only the pertinent excerpts,
with line numbers adjusted to correspond to \autoref{fig:fix-ensure}):

\vspace{1px}
\begin{mdframed}[linewidth=1pt]
  \begin{minipage}{\linewidth}
    \footnotesize
\begin{verbatim}
error: post-condition does not hold
|
|    ensures global<ObjectCore>(result.self)
|            .guid_creation_num == INIT_GUID_CREATION_NUM
| =  at  create_object_internal (Line 12)
| =      guid_creation_num = 1125899906842624
| =  at  guid.move:create (Line 26)
| =      guid_creation_num = 1125899906842625
\end{verbatim}
  \end{minipage}
\end{mdframed}

This diagnostic clearly identifies the semantic inconsistency:
the difference between \cc{guid\_creation\_num}'s value
at Line 12 (representing the pre-increment state)
and Line 26 (reflecting the post-increment state),
with the latter serving as the parameter for \cc{ObjectCore} construction.
Through \sys's feedback incorporation mechanism,
this verification failure is transmitted to the LLM,
which demonstrates remarkable capacity for semantic reasoning
by identifying the root cause and subsequently generating
a corrected specification (Line 37)
that accurately captures the increment semantics
through an appropriately formulated \cc{ensures} clause,
as corroborated by the accompanying explanatory annotations.

These examples of autonomous specification refinement
clearly demonstrate the LLM's sophisticated capacity for
formal reasoning and error correction,
establishing a feedback-driven methodology that
systematically enhances specification accuracy,
as empirically validated by the quantitative results in \autoref{tab:merged-results}.

\begin{figure*}[htb]
  \centering
  \begin{subfigure}[t]{0.48\linewidth}
    \input{code/case-modular-exp-strip.move} %
    \caption{Move implementation for natural exponential computation ($e^{x}$)}
    \label{fig:modular-abstract-spec-fun}
  \end{subfigure}
  \hfill
  \begin{subfigure}[t]{0.48\linewidth}
    \input{code/case-modular-exp.spec.move}
    \caption{Associated formal specification}
    \label{fig:modular-abstract-spec}
  \end{subfigure}
  \caption{Example of compositional and abstract specification generation by \sys.
    The \cc{exp} function implements natural exponential computation ($e^{x}$)
    through Taylor series approximation.}
  \label{fig:modular-abstract}
\end{figure*}

\PP{Compositional and Abstract Specification Generation}
Rather than constraining specification generation to monolithic constructs,
\sys incorporates teaching examples that demonstrate
modular and abstract specification methodologies,
thereby enabling the generation of more sophisticated and maintainable
formal artifacts.
\autoref{fig:modular-abstract} presents a representative example:
a function implementing natural exponential computation ($e^{x}$)
through Taylor series expansion within the aptos-stdlib framework.
The \cc{exp} function exhibits a layered architectural pattern,
delegating computational responsibilities to \cc{exp\_raw}
for high-precision arithmetic operations on \cc{u256} operands
(256-bit unsigned integers),
subsequently transforming results into \cc{FixedPoint64} representation---a
hybrid numeric type comprising distinct 64-bit integer and fractional components.
Within \cc{exp\_raw}, the implementation performs
a complex sequence of mathematical operations
encompassing multiplication, division, modular arithmetic, and bitwise manipulation.
While the underlying SMT solvers supporting Move Prover
possess robust capabilities for linear integer arithmetic,
they encounter fundamental limitations when confronted with
non-linear arithmetic constructs (exemplified by variable-to-variable multiplication),
which constitute inherent decision procedure bottlenecks.
Consequently, rather than attempting to generate
concrete specifications for \cc{exp}
(which would likely exceed current SMT solver capabilities),
\sys adopts a more nuanced approach,
producing the draft specification architecture
illustrated in \autoref{fig:modular-abstract-spec}.
This specification employs a compositional structure
comprising one primary \cc{spec} block
augmented by two auxiliary \cc{spec fun} definitions.
\cc{spec fun} constructs function as specification-level abstractions---analogous to helper functions
that undergo expansion within their associated specification contexts,
thereby facilitating modular specification design and enhanced maintainability.
This particular example shows an interesting feature:
\sys generates two \cc{spec fun} entities
(\cc{spec\_exp} and \cc{spec\_can_exp})
accompanied by comprehensive descriptive annotations
explaining their intended purposes,
yet deliberately omitting concrete definitional content.
Such constructs are formally characterized as
\emph{uninterpreted functions} \cite{uninterpreted-function}.
Uninterpreted functions serve as semantic placeholders
that abstract away implementation complexities
while preserving interface contracts---conceptually analogous to
mock objects in software testing methodologies.
Within the primary \cc{spec exp} block,
\sys employs \cc{aborts\_if [abstract]} and \cc{ensures [abstract]} annotations,
which explicitly signal to the Move Prover
that these specifications represent abstract behavioral contracts
rather than concrete verification obligations.
This example clearly shows \sys's capacity for
generating sophisticated modular and abstract specifications
that serve as high-quality drafts,
providing substantive semantic scaffolding
for subsequent expert elaboration and formal completion.

\clearpage %

\subsection{Evaluation Results of \sysaio Variants for Weaker Models GPT-4o and GPT-4o-mini}
\label{app:eval-weak}
\begin{table*}[h]
  \centering
  \begin{tabular}{c *{9}{c}}
    \toprule
    & \multicolumn{3}{c}{\sysaio} & \multicolumn{3}{c}{\sysaiom} & \multicolumn{3}{c}{\sysaionaive} \\
    \cmidrule(lr){2-4} \cmidrule(lr){5-7} \cmidrule(lr){8-10}
    & o3-mini & GPT-4o & GPT-4o-mini & o3-mini & GPT-4o & GPT-4o-mini & o3-mini & GPT-4o & GPT-4o-mini \\
    \midrule
    \nth{1}   & 132 & 103 & 84  & 140 & 109 & 92  & 106 & 83 & 57  \\
    \nth{2}   & 33  & 18  & 22  & 6  &  8   &  9   & 18  & 11 & 4   \\
    \nth{3}   & 14  & 7  &  12  & 0   & 6   & 1   & 7   & 2  & 1   \\
    \nth{4}   & 8   & 6   & 3   & 0   & 6   & 1   & 7   & 2  & 1   \\
    \nth{5}   & 4   & 7  & 1   & 1   & 3   & 2   & 3   & 2  & 2   \\
    \midrule\midrule
    Fail         & 54 & 95 & 221 & 77 & 70 & 240 & 220 & 254 & 291 \\
    Success      & 191 & 141 & 122 & 147 & 132 & 104 & 137 & 103 & 66  \\
    $ > \nth{1}$ & 59  & 38  & 38  & 7 & 23  & 12  & 31  & 20  & 9  \\
    Abstract     & 112  & 121  & 14  & 133  & 155  & 13  & 0  & 0  & 0  \\
    \midrule\midrule
    $\frac{\text{Success}}{Total}$ & \textbf{53.5\%} & 39.5\% & 34.2\% & 41.1\% & 37.0\% & 29.1\% & 38.3\% & 28.8\% & 18.4\% \\
    $\frac{\nth{1}}{\text{Success}}$       & 69.1\% & 73.0\% & 68.9\% & 95.2\% & 82.6\% & 88.5\% & 77.4\% & 80.6\% & 86.4\% \\
    $\frac{> \nth{1}}{\text{Success}}$       & \textbf{30.9\%} & \textbf{27.0\%} & \textbf{31.1\%} & 4.8\% & 17.4\% & 11.5\% & 22.6\% & 19.4\% & 13.6\% \\
    \midrule
    $\frac{\text{Success}}{\text{Success} + \text{Abstract}}$       & 63\% & 53.9\% & 89.7\% & 52.5\% & 46\% & 88.9\% & 100\% & 100\% & 100\% \\
    \bottomrule
  \end{tabular}
  \caption{Result and round detail for different variants of \sysaio:
    \sysaiom, \sysaionaive.
    \nth{1} to \nth{5} means the earliest round
    that \sysaio could
    generate the correct specification.
    The center rows show the summary of the results.
    ``Fail'' means that the
    \sysaio cannot generate the specification  within the five rounds.
    ``Success'' is the summation of \nth{1} to \nth{5}, which means that
    the \sysaio can generate the specification within five rounds.
    ``$> \nth{1}$'' is the summation of \nth{2} to \nth{5}, which means that the
    \sysaio can generate the specification within five rounds but not
    in the first
    round (i.e. LLM cannot one-shot).
    ``Abstract'' is the number of generated abstract specifications.
    The following 3 rows are the successful rates of \sys, one-shot rates,
      and non-one-shot rates.
    The last row is the percentage of the concrete (non-abstract)
    successful generated
    specifications.
  }
  \label{tab:ablation}
\end{table*}

\begin{table}[h!]
  \centering
    \begin{tabular}{c ccc}
        \toprule
        Models & o3-mini & GPT-4o & GPT-4o-mini \\
        \midrule
        \nth{1} & 132 & 109 & 93 \\
        \nth{2} & 35 & 25 & 22 \\
        \nth{3} & 15 & 6 & 9 \\
        \nth{4} & 7 & 2 & 3 \\
        \nth{5} & 4 & 7 & 3 \\
        \midrule
        \midrule
        Fail & 47 & 81 & 214 \\
        Success & 193 & 149 & 130 \\
        $> \nth{1}$ & 61 & 40 & 37 \\
        Abstract & 117 & 127 & 13 \\
    \midrule\midrule
    $\frac{\text{Success}}{Total}$ & \textbf{54.1\%} & 41.7\% & 36.4\%  \\
    $\frac{\nth{1}}{\text{Success}}$  & 68.4\% & 73.2\% & 71.6\% \\
    $\frac{> \nth{1}}{\text{Success}}$  & 31.6\% & 26.8\% & 28.4\% \\
    \midrule
    $\frac{\text{Success}}{\text{Success} + \text{Abstract}}$  & 62.3\% & 54.0\% & 91.0\% \\
      \bottomrule
    \end{tabular}
  \caption{Result and round detail of \sysaioinline. Please refer to
  \autoref{tab:merged-results} for the definition of each column. }
  \label{tab:inline}
\end{table}

\fi

\end{document}